\documentclass[a4paper,11pt]{article}
\pdfoutput=1 

\usepackage{jcappub} 

\usepackage[T1]{fontenc} 
\usepackage{graphicx}
\usepackage{amsmath}
\newcommand{\angstrom}{\textup{\AA}}
\newcommand{\hmpc}{{\, h^{-1}\, {\rm Mpc}}}

\def\aj{AJ}
\def\apj{ApJ}

\def\mnras{MNRAS}

\def\nat{Nature}

\title {\boldmath On the origin of red spirals: Does assembly bias
  play a role?}

\author[a]{Suman Sarkar,}

\author[b]{Biswajit Pandey,}

\author[b]{and Apashanka Das}

\affiliation[a]{Department of Physics, Indian Institute of Science Education and Research Tirupati, Tirupati - 517507, Andhra Pradesh, India}
\affiliation[b]{Department of Physics, Visva-Bharati University, Santiniketan, 731235, India}

\emailAdd{suman2reach@gmail.com} \emailAdd{biswap@visva-bharati.ac.in} \emailAdd{a.das.cosmo@gmail.com} 

\abstract{The formation of the red spirals is a puzzling issue in the
  standard picture of galaxy formation and evolution. Most studies
  attribute the colour of the red spirals to different environmental
  effects. We analyze a volume limited sample from the SDSS to study
  the roles of small-scale and large-scale environments on the colour
  of spiral galaxies. We compare the star formation rate, stellar age
  and stellar mass distributions of the red and blue spirals and find
  statistically significant differences between them at $99.9\%$
  confidence level. The red spirals inhabit significantly denser
  regions than the blue spirals, explaining some of the observed
  differences in their physical properties. However, the differences
  persist in all types of environments, indicating that the local
  density alone is not sufficient to explain the origin of the red
  spirals. Using an information theoretic framework, we find a small
  but non-zero mutual information between the colour of spiral
  galaxies and their large-scale environment that are statistically
  significant ($99.9\%$ confidence level) throughout the entire length
  scale probed. Such correlations between the colour and the
  large-scale environment of spiral galaxies may result from the
  assembly bias. Thus both the local environment and the assembly bias
  may play essential roles in forming the red spirals. The spiral
  galaxies may have different assembly history across all types of
  environments. We propose a picture where the differences in the
  assembly history may produce spiral galaxies with different cold gas
  content. Such a difference would make some spirals more susceptible
  to quenching. In all environments, the spirals with high cold gas
  content could delay the quenching and maintain a blue colour,
  whereas the spirals with low cold gas fractions would be easily
  quenched and become red.}

\begin{document}
\maketitle
\flushbottom


\section{Introduction}
Understanding the formation and evolution of galaxies remains one of
the most challenging goals in cosmology. The galaxies are the building
blocks of the large-scale structures in the Universe. They are
distributed along an interconnected filamentary network surrounded by
nearly empty regions. They form and evolve in different environments
of the cosmic web \citep{bond96} and they can have various shapes,
sizes, masses, colours, star formation rates (SFR) and
metallicities. The galaxies can be primarily divided into two distinct
morphological types based on their visual appearences, namely the
spirals and the ellipticals. The morphological bimodality
\citep{hubble26} in the galaxy population has been known for nearly a
century. More recently, similar bimodalities have also been observed
in galaxy colour \citep{strateva01, hogg03, balogh04, baldry04}, star
formation rate, stellar age \citep{kauffmann03a}, bulge to disc ratio
\citep{kauffmann03a} and gas to stellar mass ratio
\citep{kannappan04}. The observed bimodality in optical colour
\citep{strateva01, hogg03, kauffmann03a, blanton03, balogh04,
  baldry04} indicates that the galaxies in the present Universe can
also be segregated into two distinct populations, namely the ``blue
cloud'' and the ``red sequence''. The colour bimodality is strongly
correlated with morphological bimodality. The spirals are
predominantly found in the blue cloud, and the ellipticals are mostly
found in the red sequence.

The colour represents the stellar population in a galaxy. The blue
colour indicates active star formation, whereas the red colour is
known to be associated with quenching of star formation and an older
stellar population. The strong correlation between morphology and
colour of galaxies indicates that quenching of star formation in
galaxies are usually accompanied by a morphological transformation.
However, this may not be necessarily true. The transformation to
early-type morphologies is not required for quenching. Exception to
this correlation is now quite evident from the observational fact that
a significant number of ellipticals are part of the blue cloud
\citep{schawinski09a} and a large number of spirals are the members of
the red sequence \citep{masters10, fraser16}. Such exceptions may hold
important clues about galaxy formation and evolution. These deviations
have drawn considerable interest in recent years, and several works
have addressed this issue and tried to accommodate these findings
within the existing paradigm.

In the present work, we mainly focus on the red spirals and their
origin. The presence of the red spirals is first noted by van den
Bergh \citep{vanden} in the Virgo cluster. The red spirals are also
detected in distant clusters \citep{couch98, dressler99, poggianti99}
and clusters at low redshift \citep{goto03, moran06, wolf09,
  gallazzi09}. Although the existence of the red spirals is known for
quite some time, the data available from these observations are not
sufficient to address the issue statistically. The Galaxy Zoo project
\citep{lintott08} provides the visual morphological classifications of
nearly one million galaxies with the help of the citizen
scientists. The studies with Galaxy Zoo data reveal that $\sim 20\%$
of the spiral galaxies are contained in the red sequence
\citep{bamford09, skibba09}.

Most of the spiral galaxies are known to be actively star
forming. They preferentially reside in the less dense environments
\citep{dressler80}. Contrary to this, the red spirals are passive, and
they exhibit a preference for the higher density environment even at
fixed stellar mass \citep{bamford09}. It suggests an environment
dependent suppression of star formation in the red spirals. It has
been suggested that mild environmental effects can change the colours
of spiral galaxies without altering their morphology. A number of
physical processes such as tidal interactions, minor mergers, thermal
evaporation \citep{cowie77}, ram pressure stripping \citep{gunn72},
galaxy harassment \citep{moore96,moore98} and strangulation
\citep{larson80, balogh00, kawata08} may play a significant role in
quenching the star formation in the red spirals. Further, the fact
that the red spirals are massive implies that mass quenching
\citep{birnboim03, dekel06, keres05, gabor10} may have a role in
curtailing star formation in these galaxies. A higher bar fraction in
the red spirals suggests that bar quenching can also play a role in
suppressing the star formation in the red spirals
\citep{masters10}. The morphological quenching may also prohibit star
formation in the red spirals by stabilizing the gas disk due to the
presence of big bulges in these galaxies \citep{martig09}. Besides,
the high angular momentum of the infalling gas may settle down the gas
on the outer part, thereby quenching the star formation in the disk
\citep{peng20}. The red colour of spiral galaxies may also arise due
to several other factors such as internal reddening by dust
\citep{driver07}, high metallicity \citep{mahajan09} and low star
formation rate \citep{cortese12}.

The existence of the red spirals is clearly at odds with the standard
picture of galaxy formation and evolution. Our current understanding
of galaxy formation and evolution may need some revision in order to
explain the origin of these unconventional objects. A number of works
is devoted to explore the origin of red spirals. Masters et
al. \citep{masters10} analyze a sample of visually selected face-on
disky red spiral galaxies from the Galaxy Zoo project and propose
multiple scenarios for the origin of the red spirals. They suggest
that the red spirals could be simply old spirals that have exhausted
their gas reservoir. A strong correlation between stellar age and
environment in different studies \citep{bundy06, cooper06, cooper09}
suggests that the galaxies start to assemble earlier at higher density
environments and have a longer time to use up their
gas. Alternatively, the red spirals could be the satellite galaxies in
massive dark matter halos where they are stripped off their gas by
strangulation \citep{masters10}. The bar instabilities can also remove
the gas from the disk by driving it inwards. Tojeiro et
al. \citep{tojeiro13} analyze data from the Galaxy Zoo project and
find that the red spirals and the blue spirals share similar star
formation histories at earlier times but depart from each other only
in the last 0.5 Gyr. Their chemical composition and dust content are
similar, indicating that the red spirals are a recent descendant of
the blue spirals. They suggest that the red spirals may represent an
evolutionary link between blue spirals and ellipticals. Mahajan et
al. \citep{mahajan20} analyze the data from the Galaxy and Mass
Assembly (GAMA) survey \citep{driver11} and conclude that the red
optical colours in the nearby spiral galaxies are a direct consequence
of some environment driven processes operating on long timescales. Hao
et al. \citep{hao19} analyze data from the SDSS DR15 (Sloan Digital
Sky Survey Data Release Fifteen) MaNGA (Apping Nearby Galaxies at APO)
observations to find that the central stellar populations in red
spirals are more similar to ellipticals than to blue spirals of
similar masses. They conclude that the red spirals can not be the
evolutionary remnants of blue spirals and are likely a product of very
gas-rich mergers above $z\sim1$. Guo et al. \citep{guo20} analyze the
SDSS data to conclude that different quenching mechanisms and galaxy
interactions may jointly cause a suppression of star formation in the
red spirals.

Although some roles of environments are indicated in most of the
studies, it is still not clear which environmental processes play the
crucial role in quenching the red spirals. There is no clear consensus
on the evolutionary pathways leading to the red
spirals. \citep{fraser18} analyze a sample of red spiral galaxies from
the 2-Micron All Sky Survey (2MASS) and conclude that no single
mechanism is responsible for quenching in red spirals and only a
mixture of different mechanisms may produce the observed red spiral
population in the present Universe. \citep{mahajan20} show that at
fixed stellar mass, the red spirals inhabit denser environments. An
increased fraction of red spirals in a higher density environment does
not ensure that environment alone is sufficient to transform the
optical colour of such galaxies. Masters et al. \citep{masters10} show
that red spirals have lower star formation rates than the blue spirals
in all environments, and there are no obvious correlations between the
environment and properties of these galaxies. They suggest that the
environment alone can not quench the star formation in red
spirals. More recently, \citep{evans} analyze the the SDSS data and
find that the fraction of star forming red galaxies is nearly
independent of their environment.

Environment plays a vital role in the formation and evolution of
galaxies. The observed morphology-density relation
\citep{hubble36,dressler80,postman84} and SFR-density relation
\citep{lewis02,gomez03,kauffmann04} show that galaxy properties are
strongly correlated with their local environment. The galaxies are
part of the large-scale coherent structures such as filaments, sheets
or clusters in the cosmic web. The influence of the environment on a
galaxy may not be limited to its local density alone. For instance,
the mass, shape and angular momentum of the dark matter halos in
N-body simulations are known to be sensitive to their large-scale
geometric environments \citep{hahn07}. Also, the clustering of dark
matter halos is known to depend on halo formation time at fixed halo
mass \citep{gao05, wechsler06, croton07, gao07, musso18,
  vakili19}. This dependence of halo clustering on the assembly
history is popularly known as the `assembly bias', which has been
extensively studied in the literature \citep{dalal08, hahn09,
  zentner14, mao18}. The existence of halo assembly bias is now well
established. Similarly, galaxies are also expected to have a wide
variety of assembly history at fixed masses. The differences in the
assembly history can influence the galaxy properties, which is known
as the galaxy assembly bias. The evidence of galaxy assembly bias in
observations is highly debated in the literature. Many observations do
not find any evidence of assembly bias in the galaxy distribution
\citep{zehavi11, yan13, paranjape15, lin16, sin17, alam19}. On the
other hand, several studies reported a clear evidence of assembly bias
\citep{miyatake16, montero17, kerscher18}.

The assembly history of galaxies are correlated with their large-scale
environment. The large-scale environmental dependence of galaxy
properties may thus signal the existence of galaxy assembly bias. A
large number of studies with different observational data sets
indicate statistically significant correlations between the
large-scale environment and different galaxy properties
\citep{pandey06, pandey08, scudder12, lietzen12, darvish14, filho15,
  luparello15, pandey17, pandey20a, sarkar20, bhattacharjee20}.

In this work, we intend to investigate if the galaxy assembly bias
plays any role in the formation and evolution of red spirals. If the
assembly history of spirals plays any role in determining their colour
then the colour of these galaxies should depend on their large-scale
environment at fixed mass and morphology. Recently, Pandey \& Sarkar
\citep{pandey20a} analyze the SDSS data and find that the fraction of
red and blue galaxies depend on their geometric environment at fixed
density. The correlation between any galaxy property and the
large-scale environment can also be measured using the mutual
information between them \citep{pandey17}. Sarkar \& Pandey
\citep{sarkar20} propose an information theoretic framework to assess
the statistical significance of any observed non-zero mutual
information. We plan to use this method to probe if the optical colour
of galaxies at fixed stellar mass and fixed morphology is sensitive to
their large-scale environment. It would help us verify the role of
galaxy assembly bias in deciding the colour of spiral galaxies.

The SDSS measure the photometric and spectroscopic information of
millions of galaxies in the nearby Universe. The Galaxy Zoo
\citep{lintott08, willett13} provides the morphological classification
of nearly one million galaxies from the SDSS. They together provide an
unprecedented opportunity to study the origin of the red spirals. We
use these datasets for the analysis presented in this paper.

The plan of the paper is as follows. We present the data in Section 2,
describe the method of analysis in Section 3, discuss the results in
Section 4 and present our conclusions in section 5.


\section{SDSS Data}
\label{sec:data}

We use data from the $16^{th}$ data release \cite{ahumada20} of the
Sloan Digital Sky Survey (SDSS) \cite{york00}. The SDSS is the largest
and one of the most successful redshift surveys to date. The technical
details of the SDSS photometric camera are described in Gunn et
al. \citep{gunn98}. Gunn et al. \citep{gunn06} describe the construction,
design and performance of the SDSS telescope. The detailed algorithm
for selecting the SDSS main sample for spectroscopy is provided in
Strauss et al. \citep{strauss02}.

The SDSS, in its fourth phase, targets nearly three million galaxies
covering a vast area of $14,555$ square degrees of the sky. DR16, the
final data release of phase IV, is a superset of all the prior data
releases of SDSS to date. SDSS IV provides corrected data for previous
bad plates and includes several new objects as targets. We download
the data from the SDSS {\it
  CASjobs}\footnote{https://skyserver.sdss.org/casjobs/} using
\textit{Structured Query Language}(SQL). We consider all the objects
of class $galaxy$ with $zwarning=0$ and the apparent r-band Petrosian
magnitude $m_r<17.77$ within redshift range $0 \leq z \leq 0.2$. We
join six different tables of the DR16 database to obtain the required
information of the galaxies. The photometric and spectroscopic
information of the galaxies are taken from {\it SpecPhotoAll} and {\it
  Photoz}. The {\it galSpecIndx} table provides $4000 \angstrom$ break
strength \cite{bruzual93} derived from the MPA-JHU spectroscopic
catalogue \cite{brinch04} of galaxies. The stellar mass and star
formation rate of galaxies estimated using stellar population
synthesis model \cite{conroy09} are provided in {\it
  stellarMassFSPSGranWideDust}. The internal dust extinction of the
source using {\it Gas AND Absorption Line Fitting} (GANDALF)
\cite{sarzi06} is obtained from {\it emissionLinesPort}
\cite{Capp04,maras11,thomas11}. Finally, the morphologies are
specified in the {\it zooSpec} table, which provides the visual
classification of SDSS galaxies performed through the Galaxy Zoo
project \cite{lintott08, lintott11}. We identify elliptical and spiral
galaxies as those which have their \textit{elliptical} and
\textit{spiral} flag set to 1 (debiased vote fraction > 0.8)
respectively. We obtain this information for a total $619007$ galaxies
by combining the six tables mentioned here.

We then construct a volume limited sample using the downloaded
data. We identify a contiguous region within $ 135^{\circ} \leq \alpha
\leq 225^{\circ}$, $0^{\circ} \leq \delta \leq 60^{\circ}$ and apply a
cut in the r-band absolute magnitude $M_r\le -21$. These cuts provide
us with a volume limited sample that contains $124911$ galaxies within
$z \leq 0.12$, out of which $46261$ are spirals, $12772$ are
ellipticals and $65878$ are galaxies with uncertain morphology.

According to the need of the present work, we only retain the $46261$
spiral galaxies in our volume limited sample. We classify these
galaxies into red, green and blue classes using the technique
discussed in \autoref{sec:fuzzy}. This classification scheme divides
the galaxies according to their dust corrected $u-r$ colour. We find
that the number of blue, green and red spirals in our volume limited
samples are $29592$, $8120$ and $8549$, respectively.

One of the important aims of this work is to test the correlation
between the colour of spiral galaxies and their large-scale
environment. We need to subdivide the region occupied by the red and
blue spirals into regular cubic voxels and randomly shuffle them in
order to asses the statistical signficance of any observed
correlations between colour and large-scale environment of the
spirals. So we extract the largest cube with sides of $181 \hmpc$ that
can fit within our volume limited sample. It contains $12589$ spiral
galaxies, of which $8522$, $2133$ and $1934$ are blue, green and red,
respectively. This analysis aims to test the effects of the assembly
bias on the colour of spiral galaxies. The assembly bias manifests as
a dependence of clustering on the formation time or history at a fixed
stellar mass.  We choose a fixed stellar mass range $2\times 10^{10}
\leq \frac{M}{M_{\odot}} \leq 2 \times 10^{11}$ and apply this cut to
the galaxies in the cubic region.  After adopting the stellar mass
cut, we are left with $11800$ spiral galaxies of which $8068$ are
blue, $1963$ are green and $1769$ are red.

In this work, we use $\Lambda$CDM cosmological model with
$\Omega_{m0}=0.315, \Omega_{\Lambda 0}=0.685$ and $h=0.674$
\citep{planck18}.

\begin{figure*}[htbp!]
\centering
\includegraphics[width=14cm]{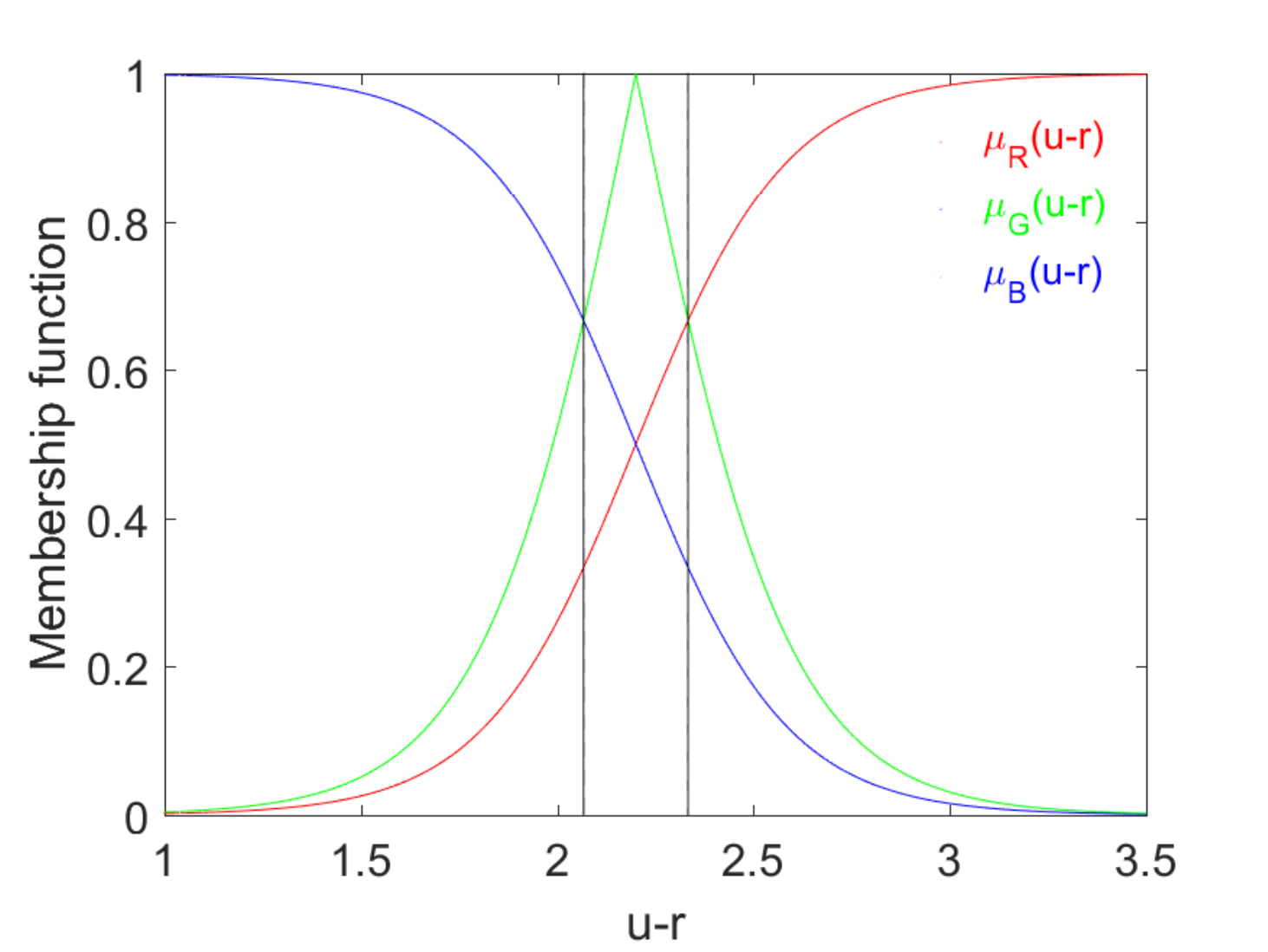}
\caption{The above figure shows the membership function for the red,
  blue and green galaxies. The region in between the two solid black
  vertical lines represents the green galaxies. The left and the right
  side of this region corresponds to the blue and the red galaxies
  respectively.}
\label{fuzzyfig}
\end{figure*}
\section{Method of analysis}
\subsection{Classifying the spirals according to their colours}
\label{sec:fuzzy}
We adopt a fuzzy set theory based classification scheme proposed in 
\citep{pandey20b}. We briefly describe the primary steps involved
in this classification.

Let us consider a fuzzy set $A$ defined as,
\citep{zadeh},
\begin{eqnarray}
A=\big{\{}\,(\,x,\,\mu_{A}(x)\,) \,\,|\,\,x \in X\,\big{\}} 
\label{eq:fuzzy1}
\end{eqnarray}
where $A$ is a set of ordered pairs $x$ and the corresponding
membership function $\mu_{A}(x)$. The membership function measures the
degree of membership of $x$ by mapping it to a real number in the
interval $[0,1]$.

This method of classification is solely motivated by the observed
bimodal distribution of galaxy colours. Our primary goal is to
separate the red spirals, and blue spirals from the volume limited
sample of spiral galaxies described in the previous section. Both a
reduced star formation activity and the presence of dust can redden
the colour of spirals. Besides correcting for the dust
  within our own galaxy, we also make corrections for the dust
  attenuation in the source galaxy. We use the internal reddening
  E(B-V) for each galaxy to make this correction.

The fuzzy set $R$ for the redness of galaxies in the volume limited
sample is define as,
\begin{eqnarray}
R=\big{\{}\, (u-r,\, \mu_{R}(u-r))\,\,|\,\, (u-r) \in X \,\big{\}}.
\label{eq:fuzzy2}
\end{eqnarray}
Here $X$ is the Universal set of dust corrected $(u-r)$ colour. The
membership function for this fuzzy set is described as
\citep{pandey20b},
\begin{eqnarray}
\mu_{R}(u-r;a,c)=\frac{1}{1+e^{-a[(u-r)-c]}}
\label{eq:fuzzy3}
\end{eqnarray}
, where $c$ and $a$ are constants representing the crossover point of
the fuzzy set and the slope at the crossover. The choice of the
sigmoidal membership function is motivated by the bimodal nature of
the $u-r$ colour distribution. The two peaks in the bimodal $u-r$
colour distribution represent the galaxies in the `blue cloud' and
`red sequence', whereas the intermediate valley is believed to be
populated by the transitional green galaxies. The observed $u-r$
colour distribution is known to dip at $(u-r)\sim2.2$ where the two
distributions for the red and blue population meet each other. The
classification according to colour becomes most uncertain at this
point. The fuzzy set $R$ has the maximum uncertainty in its membership
function at the crossover point. A value of $c=2.2$ is chosen based on
this observation. We choose $a=5.2$ to ensure that the galaxy with the
smallest and largest $(u-r)$ colour respectively has its membership
function $0$ and $1$ in the fuzzy set $R$.  Next, we define the fuzzy
set $B$ for the `blueness' of galaxies by taking a fuzzy complement of
the set $R$. The membership function of the fuzzy set $B$ can be
written as,
\begin{eqnarray}
\mu_{B}(u-r)=1-\mu_{R}(u-r), \,  \forall (u-r) \in X
\label{eq:fuzzy4}
\end{eqnarray}
Now the fuzzy set $G$ for the `greenness' of galaxies is defined by
simply taking a fuzzy intersection of the sets $R$ and $B$. The
corresponding membership function is thus defined as,
\begin{eqnarray}
\mu_{G}(u-r)= 2 \, min\big{\{}\,
\mu_{R}(u-r),\,
\mu_{B}(u-r)\,\big{\}},\, \forall
(u-r) \in X
\label{eq:fuzzy5}
\end{eqnarray}
where $min$ in \autoref{eq:fuzzy5} represents minimum operator. The
factor $2$ in the right hand side of \autoref{eq:fuzzy5} is multiplied
so as to ensure that galaxies with $(u-r) = 2.2$ are maximally green
with $\mu_G(u-r)$ value equal to 1 (\autoref{fuzzyfig}).

Using the scheme described above, we classify red galaxies as those
for which $\mu_R(u-r)$ dominates $\mu_B(u-r)$ and $\mu_G(u-r)$. The
blue and green galaxies are also classified similarly. For the present
analysis, we find that galaxies having $(u-r)\ge 2.333$ are red,
$(u-r)\le 2.067$ are blue and $2.067 < (u-r) < 2.333$ are green
(\autoref{fuzzyfig}). We classify the $46261$ spiral galaxies in our
volume limited sample as red, blue and green based on the above
criteria. We find that our volume limited sample contains $29592$ blue
spirals, $8549$ red spirals and $8120$ green spirals.

\subsection{Comparing the properties of red and blue spirals using KS-test}
\label{sec:kstest}

We consider all the blue and red spirals ($29592$ blue and $8549$ red)
in our volume limited sample and determine the
probability distribution functions (PDFs) of their
stellar mass, star formation rate, D4000 and local density. The
stellar mass, star formation rate and D4000 for the blue and red
spirals are obtained from the SDSS database as described in
\autoref{sec:data}. We calculate the local number density for each of
the galaxies using $k^{th}$ nearest neighbour method
\citep{casertano85}. We measure the distance between a galaxy and its
$k^{th}$ closest galaxy, which we denote as $r_k$. The local number
density around a galaxy is given by
\begin{eqnarray}
{\eta}_k = \frac{k-1}{V(r_k)}  
\label{eqn:knn}
\end{eqnarray}  
where, $V(r_k)=\frac{4}{3}\pi r_k^3$. We choose $k=5$ for this analysis.\\

We want to test the differences in the properties of the blue
and red spirals. The null hypothesis assumes that the red and blue
spirals have identical probability distributions for these
properties. We test the null hypothesis using a Kolmogorov-Smirnov
test.

The Kolmogorov-Smirnov test is a non-parametric test that does not
make any assumptions about the distributions. We calculate the maximum
difference between cumulative distribution functions (CDFs) for the
two samples. The supremum difference between the two CDFs ($D_{KS}$)
is defined as
\begin{eqnarray}
D_{KS} & = & \sup_{X} \, \, \{ \,\, | f_{1,
  m}(X)-f_{2,m}(X) | \,\, \}
  \label{eqn:dks}
\end{eqnarray}
$f_{1,m}(X)$ and $f_{2,m}(X)$ are the cumulative distribution
functions of a chosen property ($X$) in the $m^{th}$ bin for the red
and blue spirals respectively. Here $m \in \{1,2,3...., N^{'} \}$ and
$\sup$ operator represents the supremum of all the
($N_{1}^{'}+N_{2}^{'}$) differences. Here $N_{1}^{'}$ and $N_{2}^{'}$
are the number of red and blue spirals in the sample.

One can test the null hypothesis at different significance level
$\alpha$ to find if the PDFs for the red and blue spirals are
significantly different. One can obtain the critical value of the
supremum difference corresponding to a given significance level
($\alpha$) as,
 \begin{eqnarray}
D_{KS} (\alpha) & = & \sqrt{- \ln \left( \frac{\alpha}{2} \right) \,\,
  \, \frac{ N_{1}^{'} + N_{2}^{'}}{2 N_{1}^{'} N_{2}^{'}}}
\label{eqn:aks}
\end{eqnarray}
 The null hypothesis may be rejected or accepted depending on whether
 the measured value $D_{KS}$ is greater or smaller than its critical
 value at a given significance level $\alpha$. If $D_{KS}>D_{KS}
 (\alpha)$, then the null hypothesis is rejected at a significance
 level $\alpha$.

\subsection{Mutual information between color and environment of spirals}
\label{sec:cemi}

We extract the largest cubic region with side $L \hmpc$ that fit
within our volume limited sample and consider only the red and blue
spirals within it. We subdivide the entire cube in $N_{d}$ number of
$d \hmpc \times d \hmpc \times d \hmpc$ voxels. A discrete random
variable $X$ is defined with $N_d$ outcomes $\{ x_i:i=1,...N_d \}$
which corresponds to the environment at length scale $d$. The
probability of finding a randomly selected galaxy in the $i^{th}$ cube
is $p(x_i)=\frac{N_i}{N}$, where $N_i$ is the number of galaxies in
the $i^{th}$ voxel and $N$ is the total number of galaxies in the
cube. The Shannon entropy associated with the environment at scale $d$
is given by
\begin{eqnarray}
H(X)& = &-\sum_{i=1}^{N_d} p(x_i) \log p(x_i) \nonumber \\
&=&\log N - \frac{N_i \log N_i}{N}
  \label{eqn:Hx}
\end{eqnarray}

We use another variable $Y$ to describe the colour of the
galaxies. Our data consists of only spiral galaxies that are either
blue or red. If the cube consists of $N_{b}$ blue spirals and $N_{r}$
red spirals then the information entropy for colour will be
\begin{eqnarray}
H(Y)& = &- \left( \frac{N_{b}}{N} \log \frac{N_{b}}{N} + \frac{N_{r}}{N} \log \frac{N_{r}}{N} \right) \nonumber \\
&=& \log N- \frac{ N_{b} \log N_{b} + N_{r} \log N_{r}}{N}
  \label{eqn:Hy}
\end{eqnarray}

One can determine the mutual information between the environment ($X$)
of the galaxies and their colour ($Y$). The mutual information is
defined as
\begin{eqnarray}
I(X;Y) & = & H(X)+H(Y)-H(X,Y)
  \label{eqn:Ixy}
\end{eqnarray}

If $N_{ij}$ is the number of galaxies in the $i^{th}$ voxel that belongs to the $j^{th}$ 
colour, then $H(X,Y)$ is the joint entropy given by
\begin{eqnarray}
H(X,Y) &=& -\sum_{i=1}^{N_d} \sum_{j=1}^{2} p(x_i,y_j) \log p(x_i,y_j) \nonumber \\
&=&\log N - \frac{1}{N}\sum_{i=1}^{N_d} \sum_{j=1}^{2} N_{ij} \log N_{ij} 
\label{eqn:Hxy}
\end{eqnarray}

Where 
\begin{eqnarray}
\sum_{i=1}^{N_d} \sum_{j=1}^{2} N_{ij}=N
\label{eqn:Nij}
\end{eqnarray}

and $p(x_i,y_j)=p(x_i|y_j)p(y_j)=\frac{N_{ij}}{N}$ following Bayes'
theorem.

The two random variables may share information about each other and
$H(X,Y)$ is a measure of the information mutually shared by them. In
other words, $H(X,Y)$ is the reduction in uncertainty in one random
variable given the knowledge of the other.

\subsection{Randomizing the colour tags of blue and red spirals}
\label{sec:rtag}

We take each of the spiral galaxies in the cube and randomly tag them
as blue spirals or red spirals, obliterating their actual colour. We
do it in such a way that both the number of red spirals and blue
spirals in the new distribution remains the same as the original one.
Here the position of galaxies being unchanged, the entropy $H(X)$
would be unaltered at a given length scale. Similarly, the value of
$H(Y)$ would not change as the number of galaxies in each category
(red and blue) remains the same. However, this would change the joint
entropy $H(X,Y)$ by destroying any existing correlation between the
colour of spiral galaxies and their environment. The joint probability
distribution, in this case, would be simply a product of the
individual probabilities, $p(X_i, Y_j ) = p(X_i) p(Y_j)$. Ideally, the
randomization of colour tags should completely erase any non-zero
mutual information between the environment and the colour of the
spiral galaxies. One can test if the observed mutual information is
physical or not by comparing the mutual information $I(X;Y)$ in the
randomized and the original distributions and assessing the
statistical significance of the correlations at different length
scales.

\subsection{Shuffling of subdivided cubes}
\label{sec:shuffle}

In \autoref{sec:rtag}, we discuss how the randomization of the colour
tags would destroy any existing correlations between the colour of
spiral galaxies and their environment. One can also destroy any such
existing correlations by randomizing the spatial distribution of the
galaxies while keeping their colour tags intact. We divide the entire
cube of size $L \hmpc$ containing the spiral galaxies in smaller
sub-cubes of size $l_s=\frac{L}{n_s}$. Here $n_s$ is the number of
segments made along each side of the cube. We ensure that the side of
the sub-cubes (shuffling length) corresponding to each $n_s$ is not
equal or an integral multiple of the grid size employed for
calculation of the mutual information. We randomly pick any of the
$N_c={n_s}^3$ subcubes and allow them to swap positions. A random
rotation of the sub-cubes in multiples of $90^{\circ}$ is performed
each time. This random shuffling process followed by random rotation
is repeated $100 \times N_c$ times so that all the cubes are properly
shuffled. We carry this whole exercise for three choices $n_s=3$,
$n_s=7$ and $n_s=15$ that corresponds to shuffling lengths of $\sim 60
\hmpc$, $\sim 26\hmpc$ and $\sim 12 \hmpc$ respectively.

The shuffling of the subcubes does not affect the spiral galaxies'
colour and only alters their spatial distribution within the cube.  It
destroys all the coherent patterns in the spatial distribution above
the shuffling length $l_s$, resulting in a significant reduction in
the mutual information $I(X;Y)$. The clustering of the galaxies on a
length scale $<l_s$ would remain nearly intact. Most of the coherent
features spanning up to $l_s$ would survive the shuffling
procedure. However, the shuffling may also destroy such features if
they lie across the subcubes. So we also expect a small reduction in
$I(X;Y)$ below $l_s$. The shuffling may also introduce some spurious
spatial patterns due to pure chance alignments. However, these random
features are not expected to introduce a physical correlation between
the environment and the colour of the spiral galaxies. A comparison of
the mutual information in the shuffled and the original distributions
can be used to test the statistical significance of the observed
correlations between environment and colour.

\subsection{Testing statistical significance of the mutual information with t-test}
\label{sec:ttest}

We test the statistical significance of the mutual information between
colour and environment of the spiral galaxies using an equal variance
$t$-test. The t-test is carried out to determine if the means of two
sets of data are significantly different from each other. The $t$
score at each length scale is determined as,
\begin{eqnarray}
t= \frac{|\bar{X_1}-\bar{X_2}|}{\sigma_s \sqrt{\frac{1}{n_1}+\frac{1}{n_2}}}
\label{eqn:ttest}
\end{eqnarray}
where $\bar{X_1}$ and $\bar{X_2}$ are the average values, $\sigma_1$
and $\sigma_2$ are the standard deviations and $\sigma_s =
\sqrt{\frac{(n_1-1)\sigma_1^2+(n_2-1)\sigma_2^2}{n_1+n_2-2}}$. Here,
$n_1$ and $n_2$ are the number of datapoints associated with the two
distributions and $(n_1+n_2-2)$ is the degree of freedom in this
test. A large t-score indicates a significant difference between the
means of the two datasets whereas a small t-score indicates that the
means are similar. We use a significance level $\alpha=0.0005$ that
corresponds to $99.9\%$ confidence.

The null hypothesis assumes that the average mutual information in the
two distributions is statistically similar at any given length
scale. The randomization of colour and shuffling the spatial
distribution may reduce the mutual information statistically. A
statistically significant reduction in the mutual information conveys
a physical correlation between the two random variables.

\section{Results}

\begin{figure*}[htbp!]
\centering \includegraphics[width=7cm]{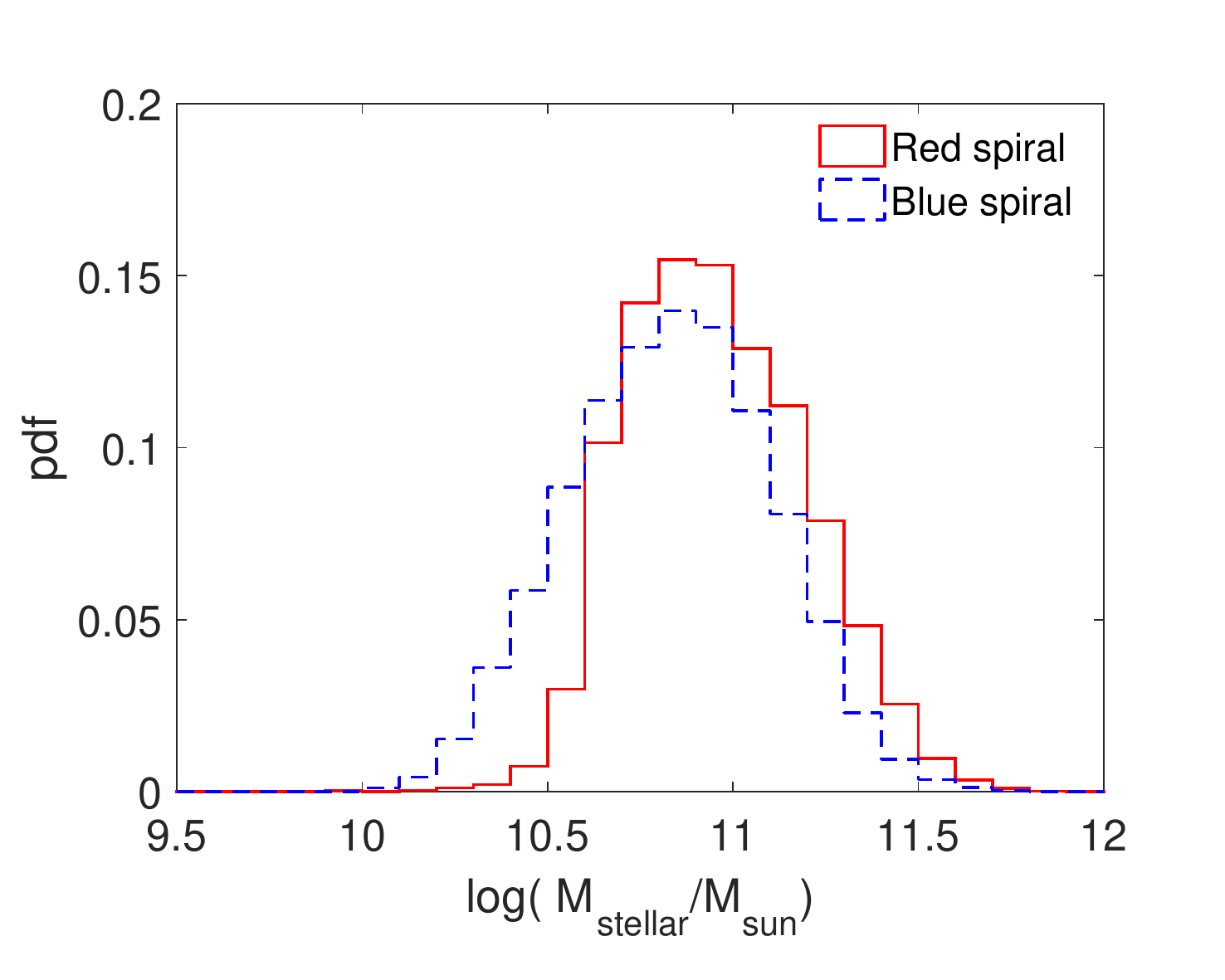}
\vspace{-0.4cm}
\includegraphics[width=7cm]{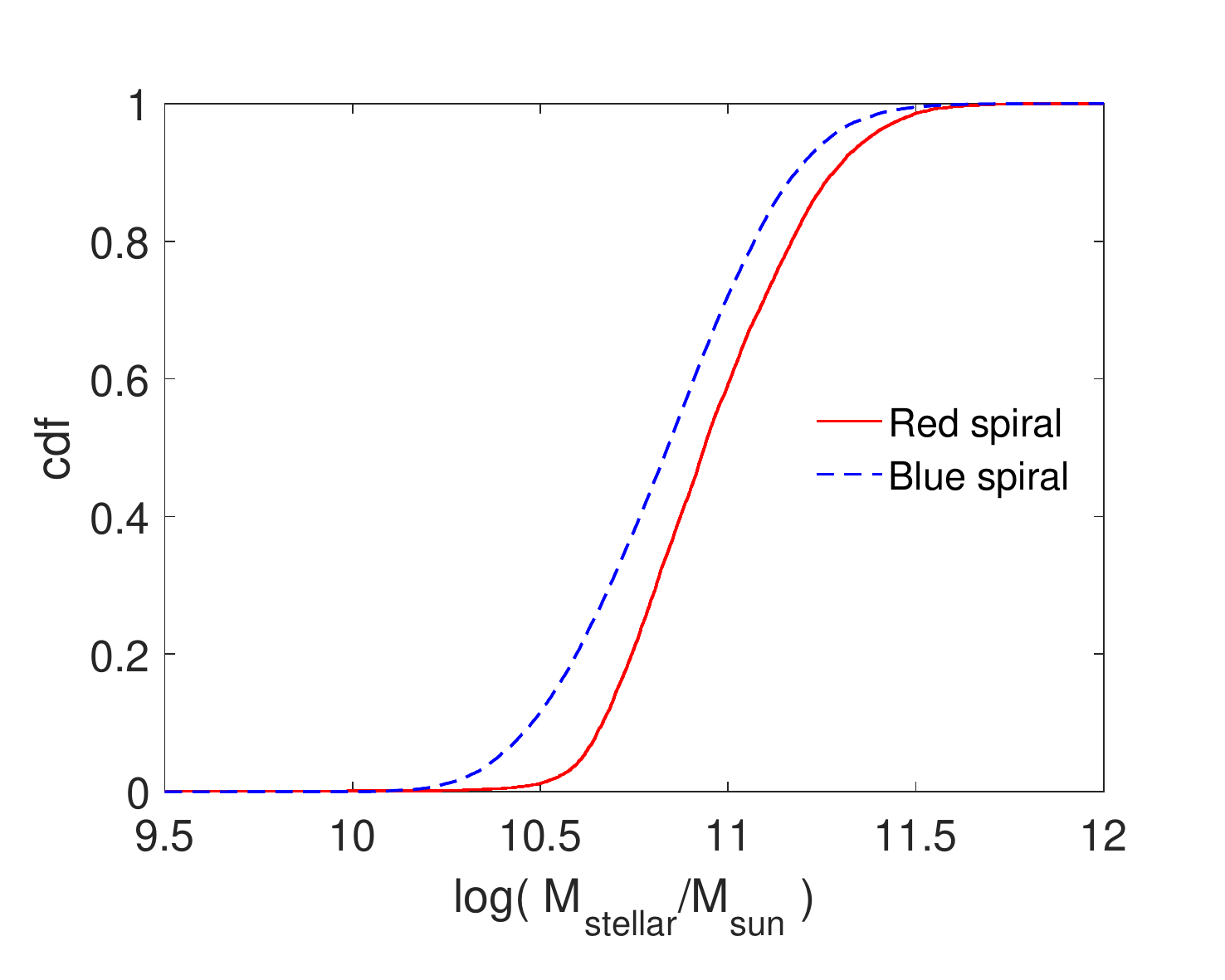}
\includegraphics[width=7cm]{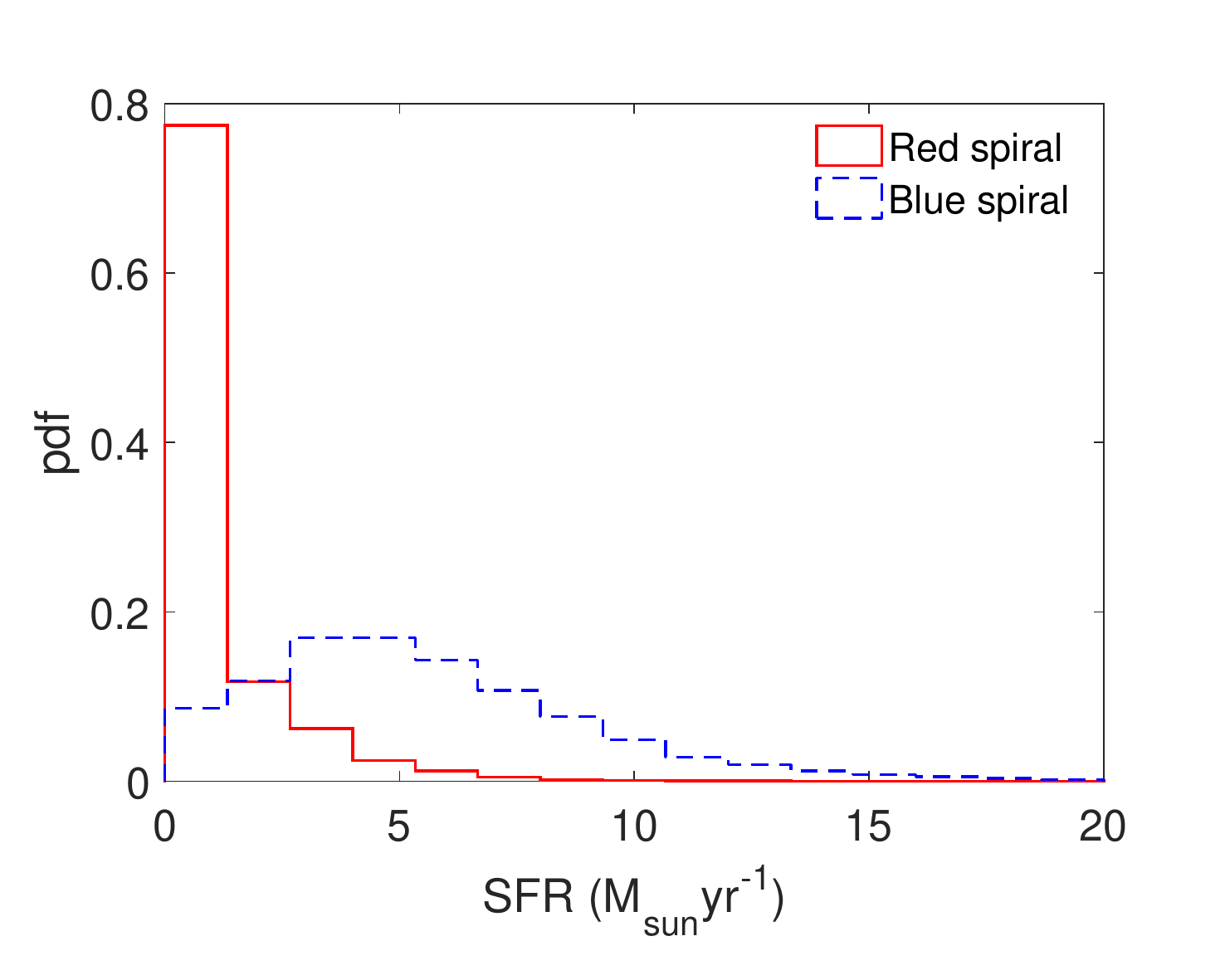}
\vspace{-0.4cm} \includegraphics[width=7cm]{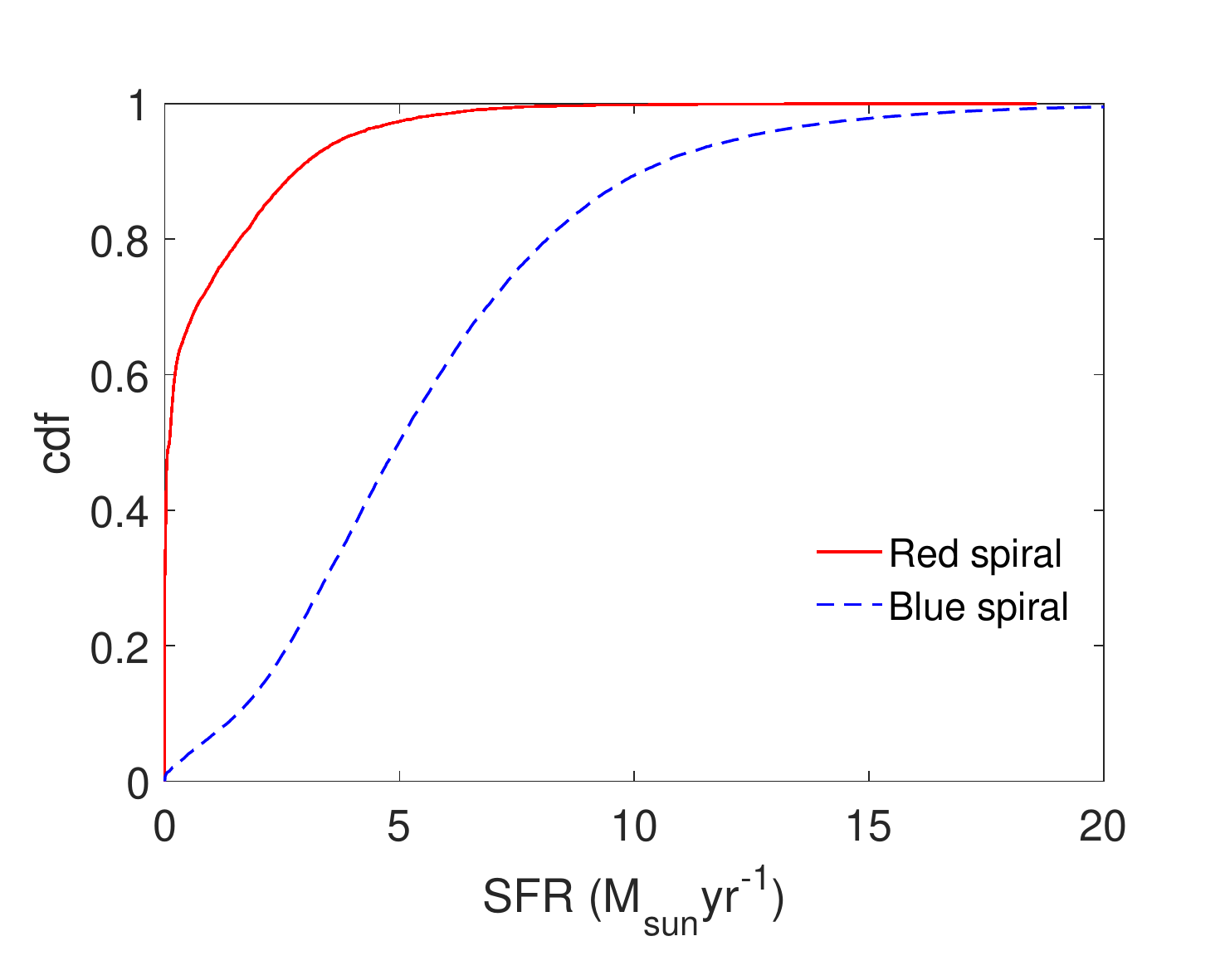}
\includegraphics[width=7cm]{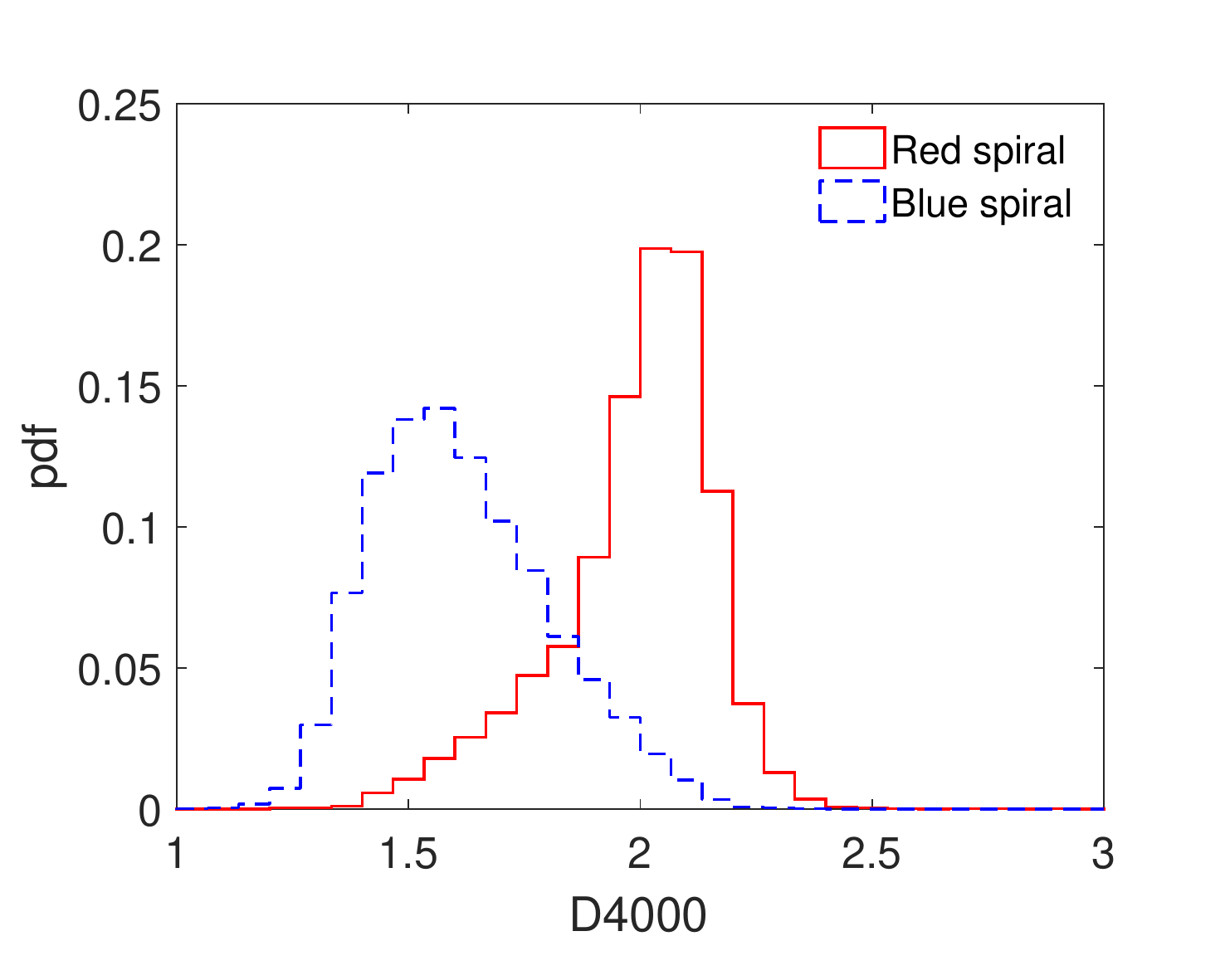}
\includegraphics[width=7cm]{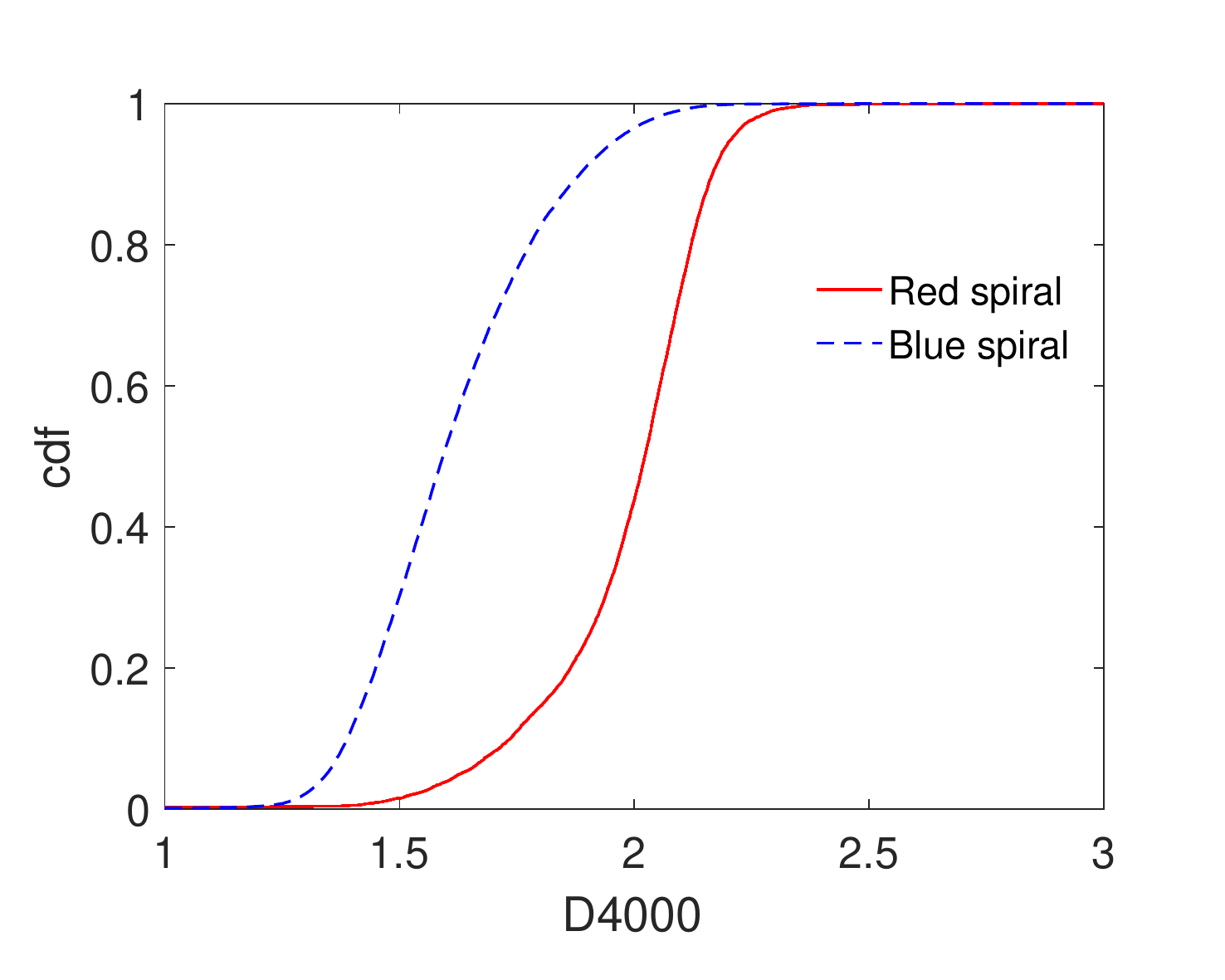}
\includegraphics[width=7cm]{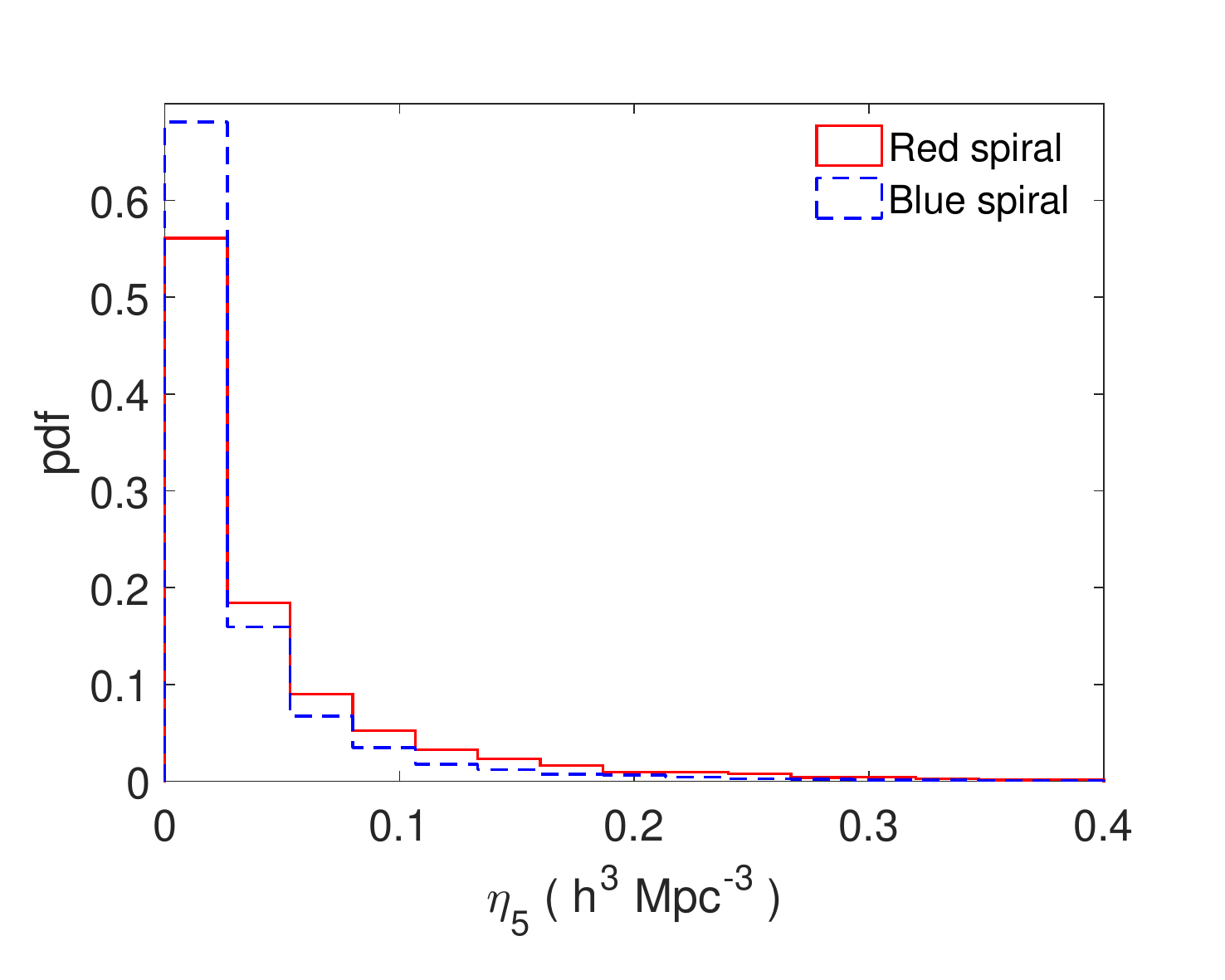}
\includegraphics[width=7cm]{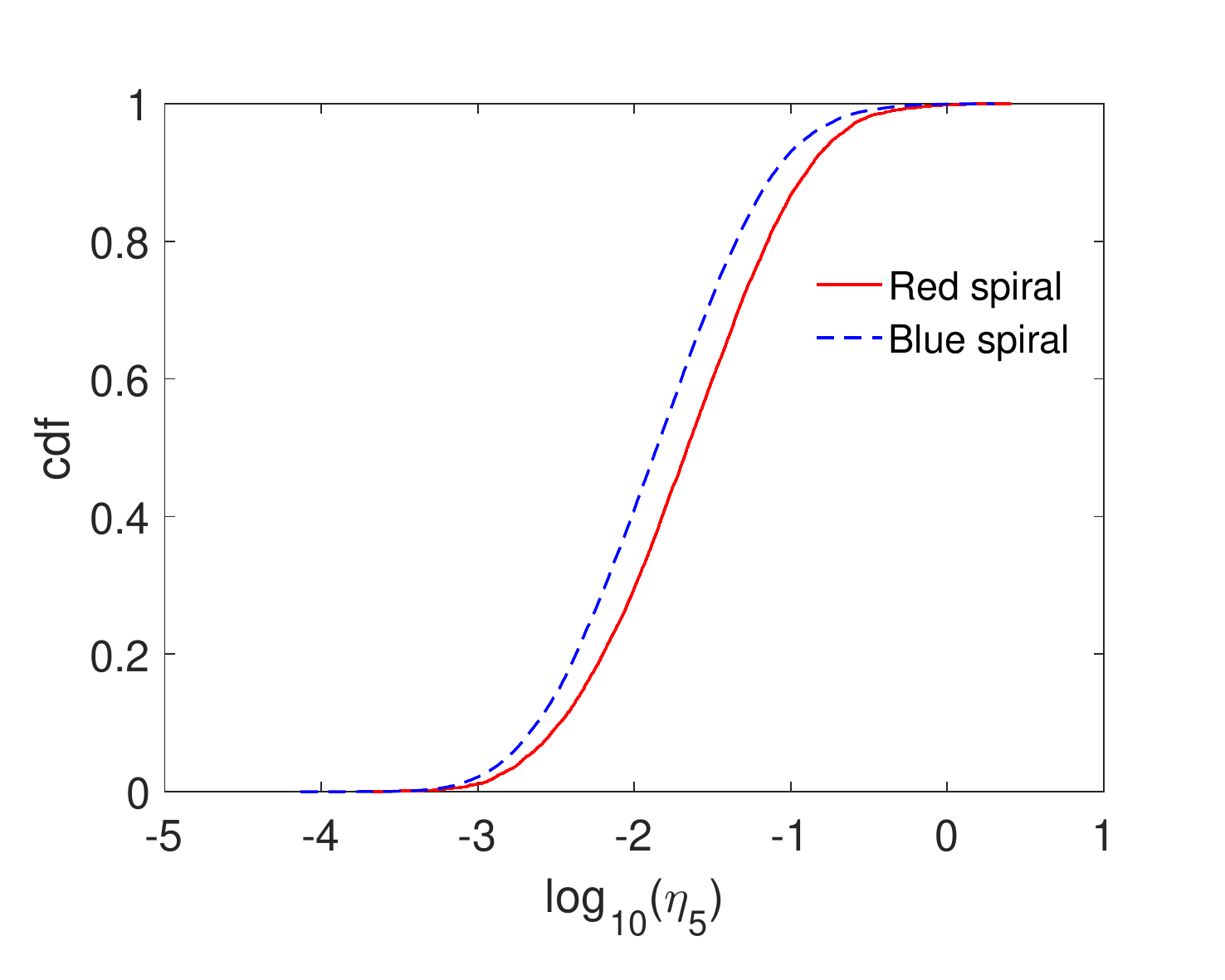}
\caption{The four left panels in this figure respectively show the
  PDFs of $\log(M_{stellar}/M_{sun})$, star formation rate (SFR), local
  density ($\eta_{5}$) and $D4000$ for the red spirals and blue
  spirals in our volume limited sample. The corresponding CDFs are
  shown in the four right panels of this figure.}
\label{fig:pdfcdf}
\end{figure*}

\begin{figure*}[htbp!]
\centering \includegraphics[width=7cm]{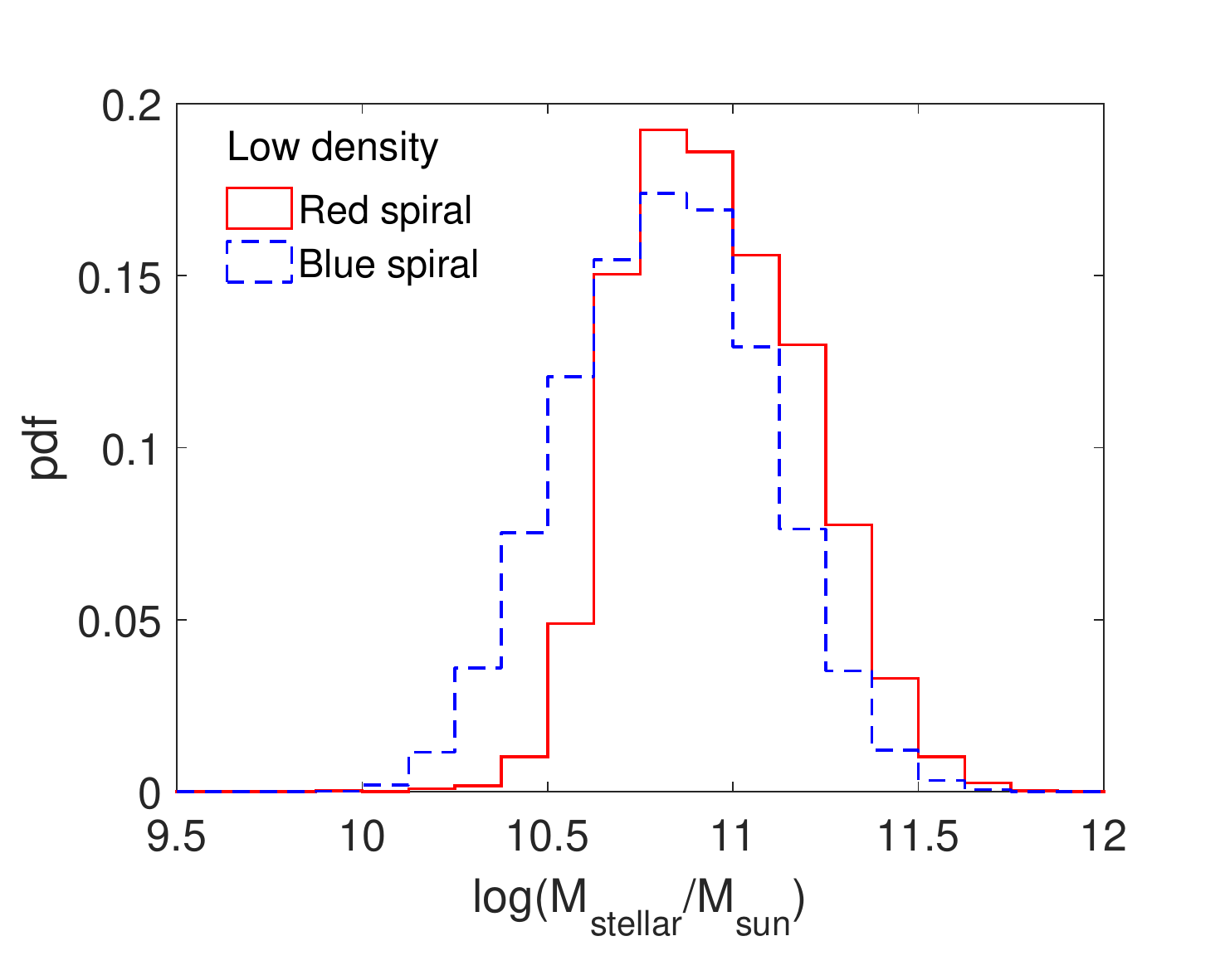}
\vspace{-0.4cm}
\includegraphics[width=7cm]{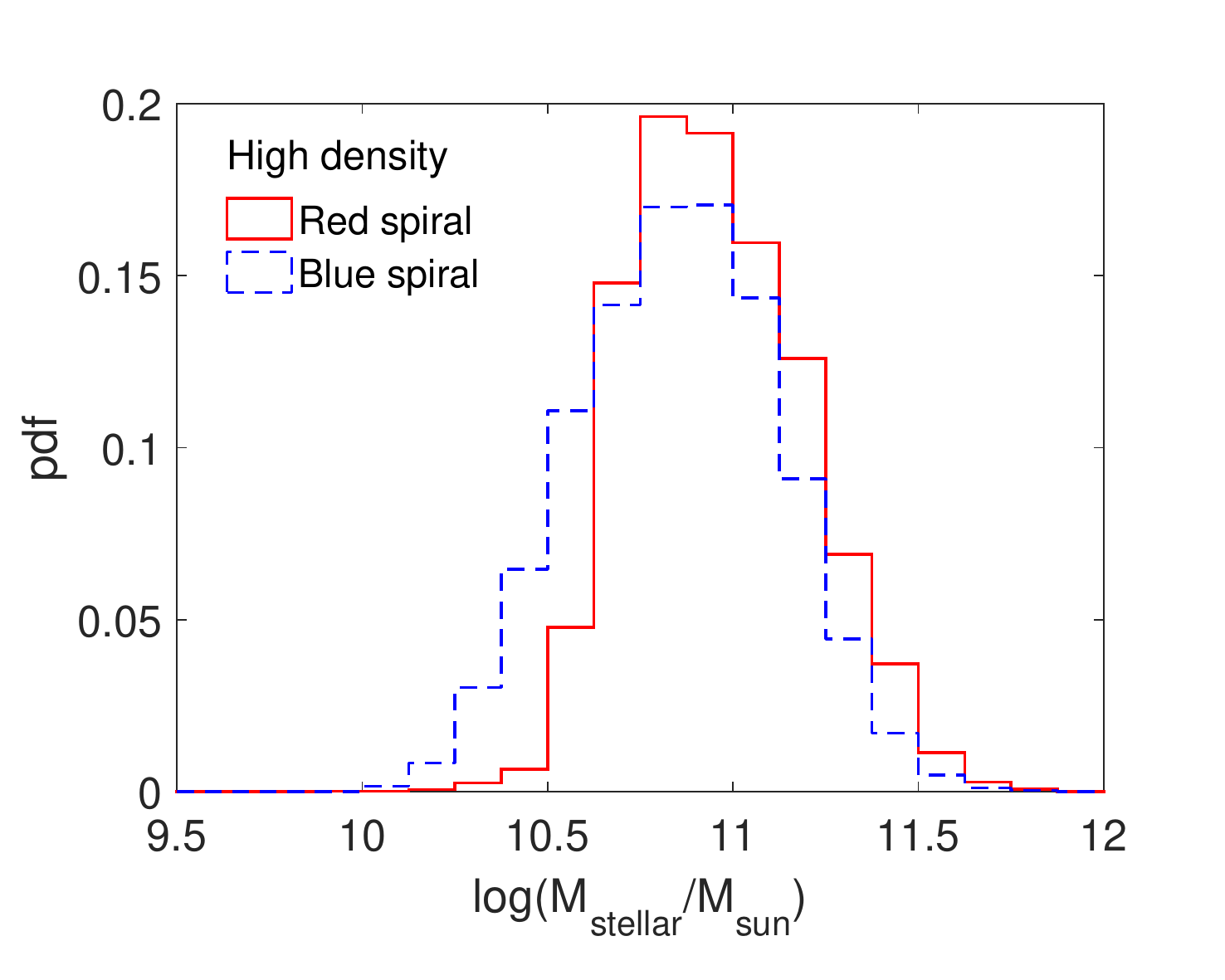}
\includegraphics[width=7cm]{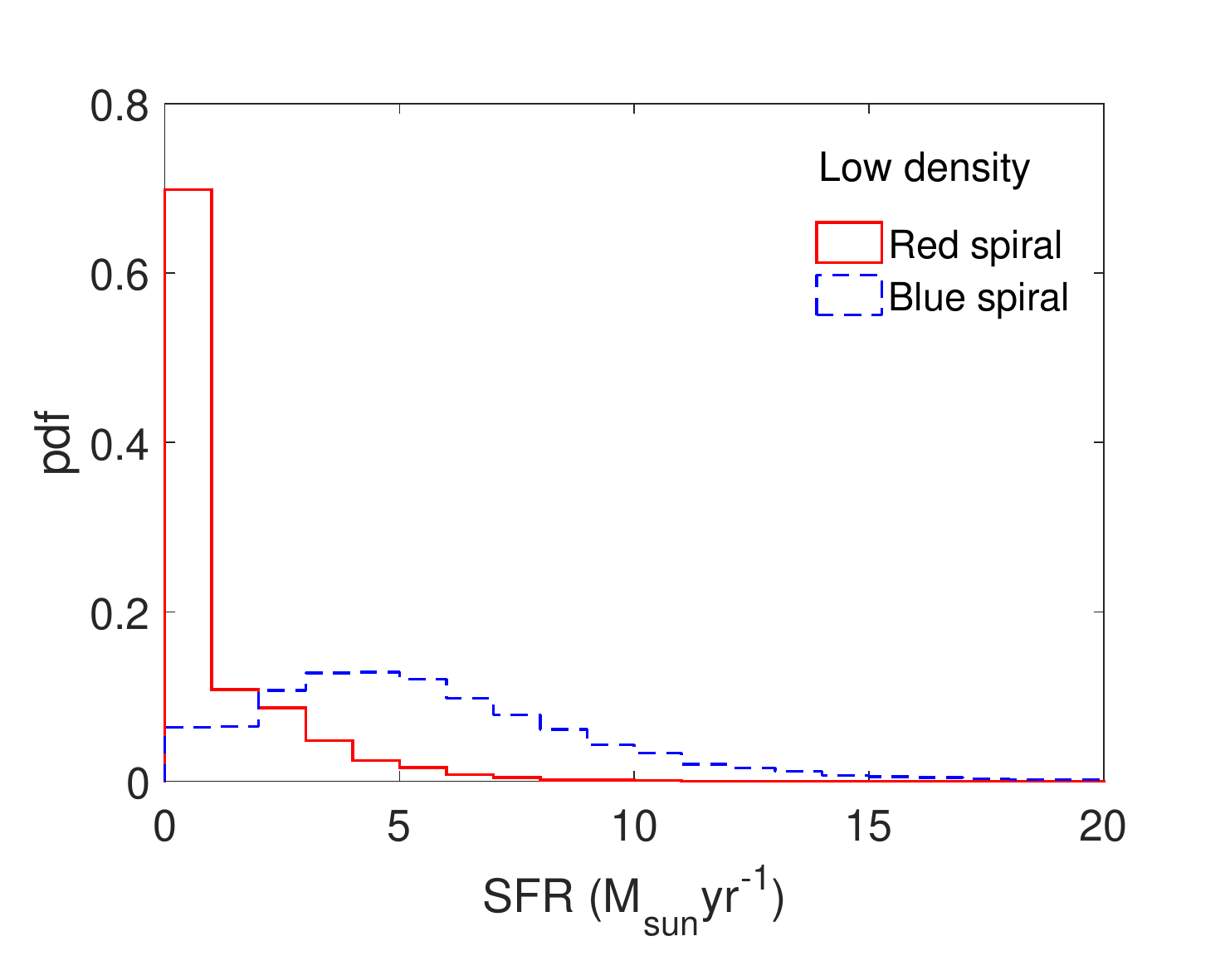}
\vspace{-0.4cm} \includegraphics[width=7cm]{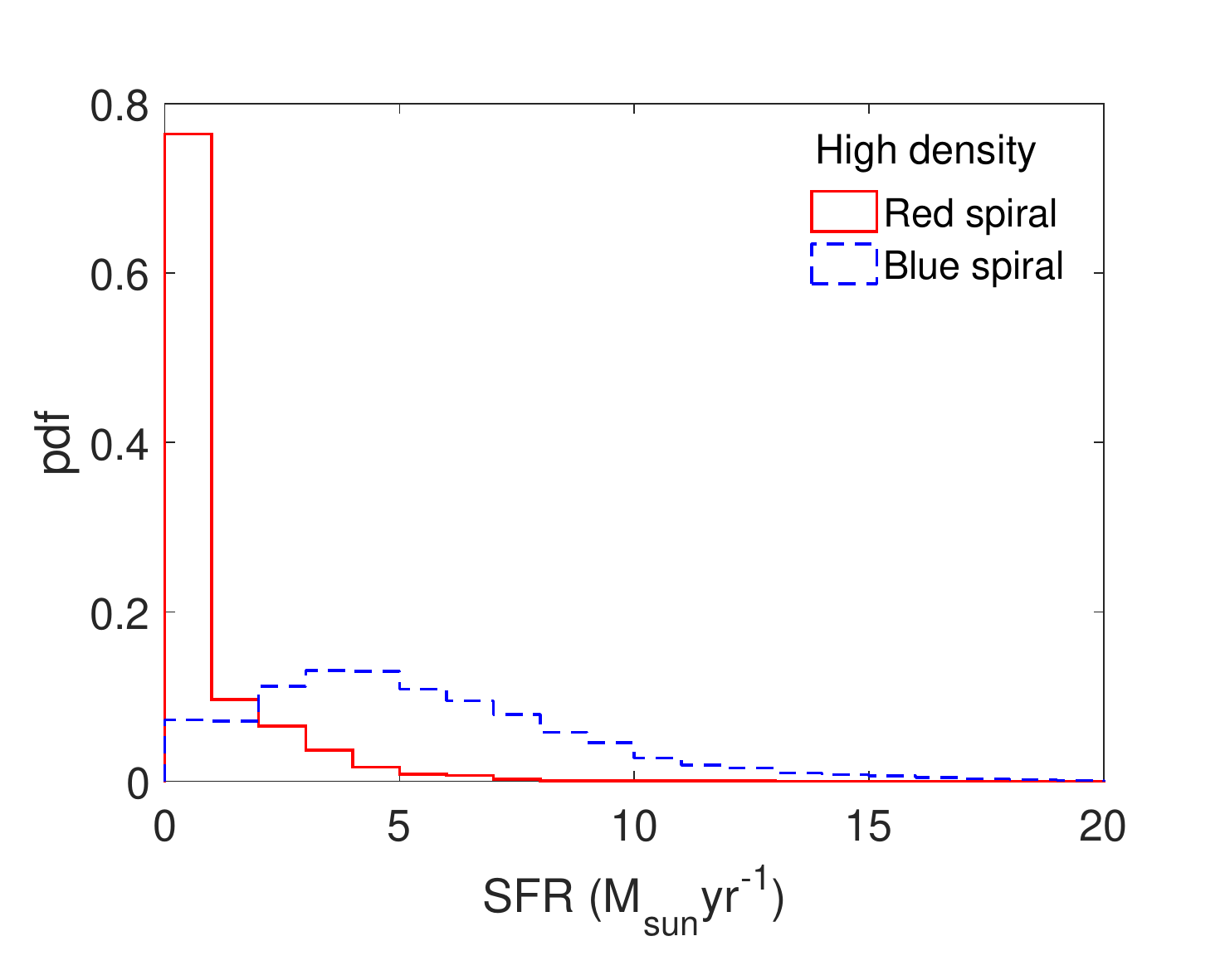}
\includegraphics[width=7cm]{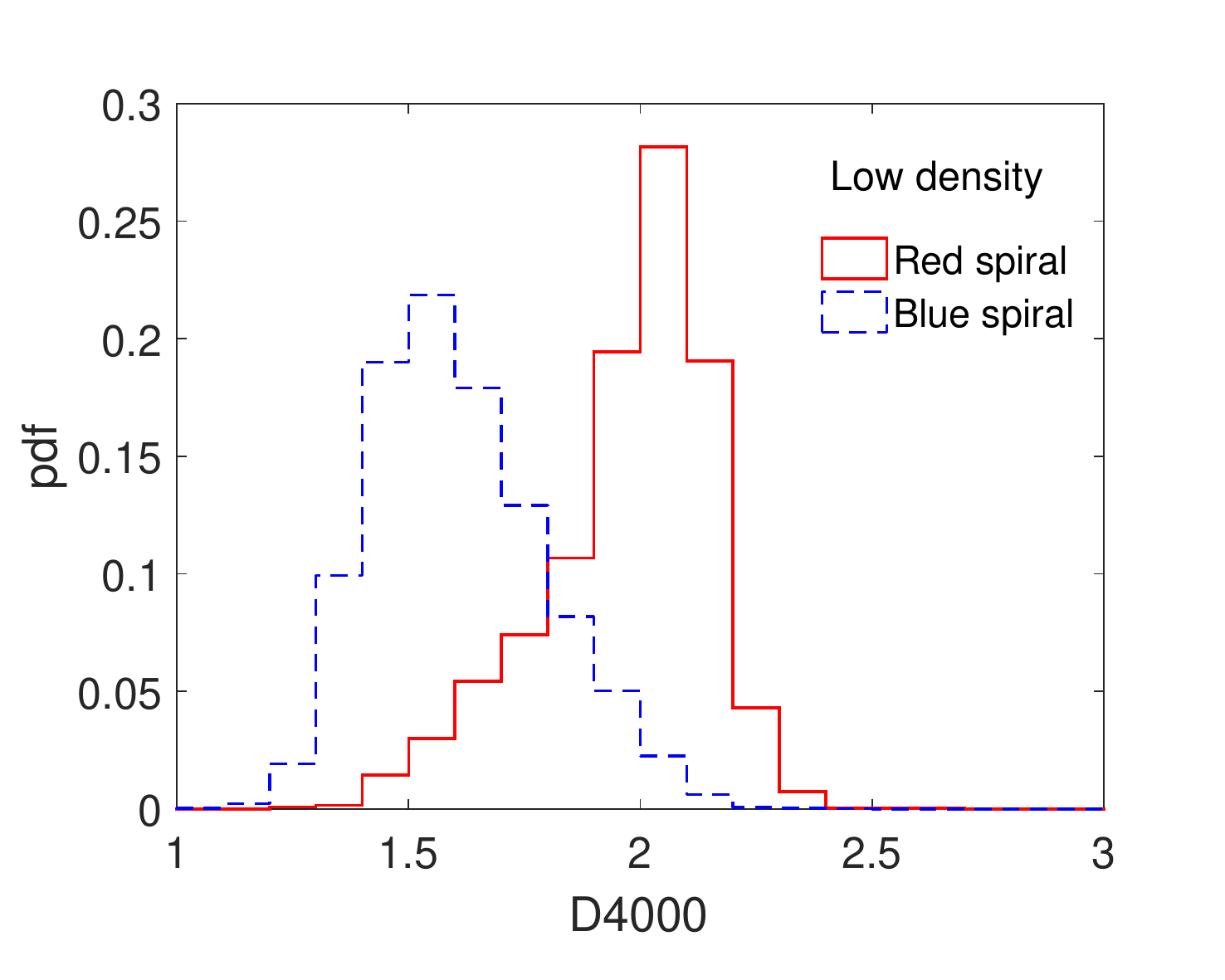}
\includegraphics[width=7cm]{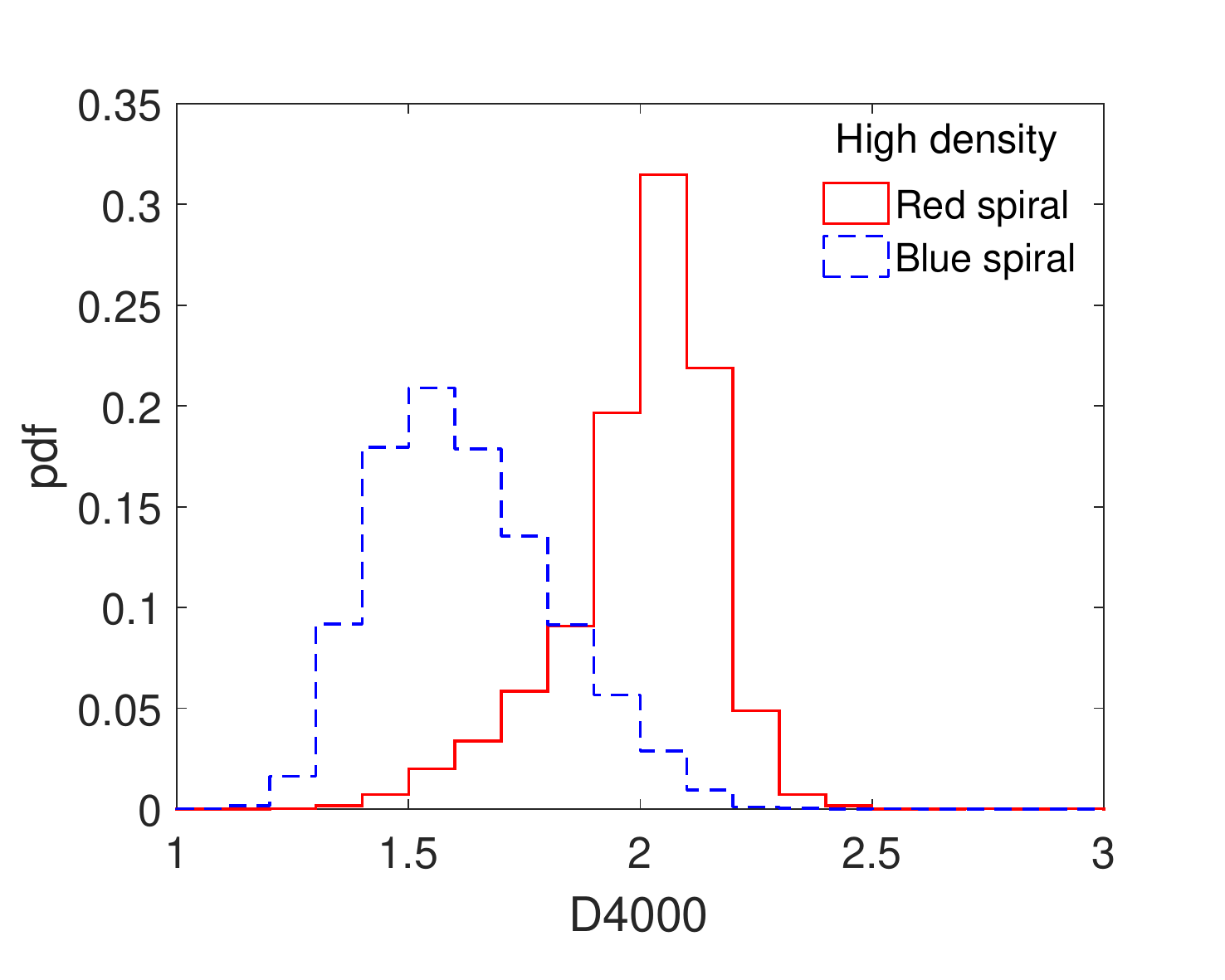}
\caption{The four left panels of this figure show the PDFs of
  $\log(M_{stellar}/M_{sun})$, star formation rate (SFR) and $D4000$
  for the red spirals and blue spirals in the low-density regions. The
  four right panels show the same in the high-density regions.}
\label{fig:lowhigh}
\end{figure*}

\subsection{Physical properties and the local environments of red and blue spirals}

We compare the PDFs of different galaxy properties and local
environments for the red and blue spirals in the left panels of
\autoref{fig:pdfcdf}. The corresponding CDFs are shown in the
respective right panels of the same figure. Different panels clearly
show that the distributions are noticeably different for the red and
the blue spirals in each case. It is clear that the distributions of
the physical properties like stellar mass, star formation rate and
$D4000$ are significantly different for the red and blue spirals. The
red spirals tend to be more massive compared to the blue spirals. They
host an older stellar population with a relatively lower star
formation rate than the blue spirals. We assess the statistical
significance of the differences in each case with a Kolmogorov-Smirnov
test. The results of these tests are tabulated in
\autoref{tab:kstab}. We find that the null hypothesis can be rejected
at $99.9\%$ confidence level for each property.  We note that the
local density of the red and blue spirals are also different in a
statistically significant way where the null hypothesis can also be
rejected at $99.9\%$ confidence level. Observations suggest that the
red spirals tend to inhabit relatively denser regions compared to the
blue spirals. So the local environment may play some role in the
formation of the red spirals. The critical question is whether these
differences in the physical properties of the red and blue spirals are
solely due to their local environments alone. It is well known that
the more massive galaxies are generally found in relatively denser
regions. But the differences in the stellar mass distributions of the
red and blue spirals may not purely arise due to the differences in
their environmental density alone. \citep{mahajan20} show that at
fixed stellar mass, the red spirals prefer denser environments. They
also find that the stellar mass of red spirals is anti-correlated with
the density of their environment. A number of other studies indicate
that there are no significant correlation between the environment and
the colour of spiral galaxies \citep{masters10, evans}.

We also test if the differences in the physical properties of the red
and blue spirals persist in different types of environments. We divide
the combined sample of red and blue spirals into two subsamples based
on the local density. The median density of the sample is used to
separate the galaxies in the high-density and low-density regions. We
define the red and blue spirals with local density above the median as
from high-density regions and those with a local density below the
median from low-density regions. There are $13987$ blue spirals and
$5084$ red spirals in the high-density regions and $15605$ blue
spirals and $3465$ red spirals in the low-density regions. Our goal is
to separately compare the PDFs of different physical properties of the
red and blue spirals in the low-density and high-density regions. The
comparisons are shown in \autoref{fig:lowhigh}. Only the PDFs of
stellar mass, star formation rate and $D4000$ for the red and blue
spirals in the low-density and high-density regions are shown in this
figure. The CDFs are not shown here. We test the differences in these
PDFs using Kolmogorov-Smirnov test in each case. The tests show that
the PDFs of different physical properties for the red and blue spirals
are statistically different at $99.9\%$ significance level both in the
low-density and high-density regions. It indicates that the local
environment alone can not explain the origin of the red spirals. The
assembly history of the spiral galaxies may also have a role in their
formation. We test this possibility using an information theoretic
framework and discuss the results in the following subsection.

\begin{table}
\centering
\begin{tabular}{|c | c | c c c c c|}
\hline
& {$D_{KS}$} & \multicolumn{5}{c|}{$D_{KS}(\alpha)$}\\

& & 99.9\% & 99\% & 90\% & 80\% & 70\% \\
\hline
$log(M_{stellar}/M_{sun})$ & 0.1791  & &  &  &  & \\
SFR & 0.7040 & 0.0239 & 0.0200 & 0.0150 & 0.0132 & 0.0120 \\
Local density & 0.1265 & & &  &  &  \\
\hline
$D4000$ & 0.0487 & 0.0240 & 0.0202 & 0.0152 & 0.0133 & 0.0121\\
\hline
\end{tabular}
\caption{The above table shows the Kolmogorov-Smirnov statistic
  $D_{KS}$ for comparison of $\log(M_{stellar}/M_{sun})$, star
  formation rate (SFR), local density ($\eta_{5}$) and $D4000$ of red
  and blue spiral galaxies. The table also lists the critical values
  $D_{KS}(\alpha)$ above which null hypothesis can be rejected at
  different confidence levels. The results for $D4000$ is shown
  separately as this information were available for $29019$ blue
  spirals and $8371$ red spirals out of total $29592$ blue and $8549$
  red spirals.}
\label{tab:kstab}
\end{table}

\subsection{Effects of randomization on the mutual information}

\begin{figure*}
\resizebox{7.5 cm}{!}{\rotatebox{0}{\includegraphics{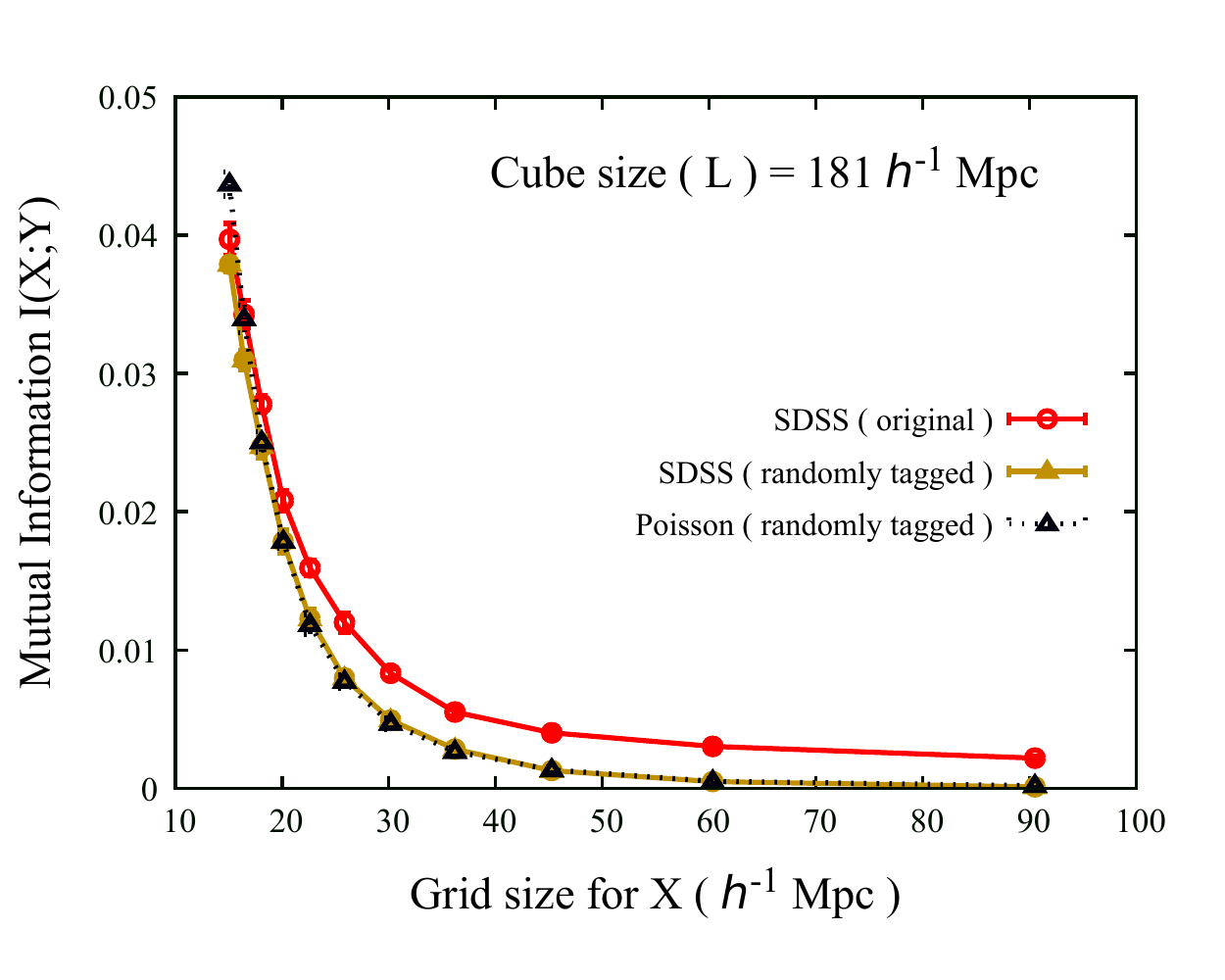}}} \hspace{0.1 cm}
\resizebox{7.5 cm}{!}{\rotatebox{0}{\includegraphics{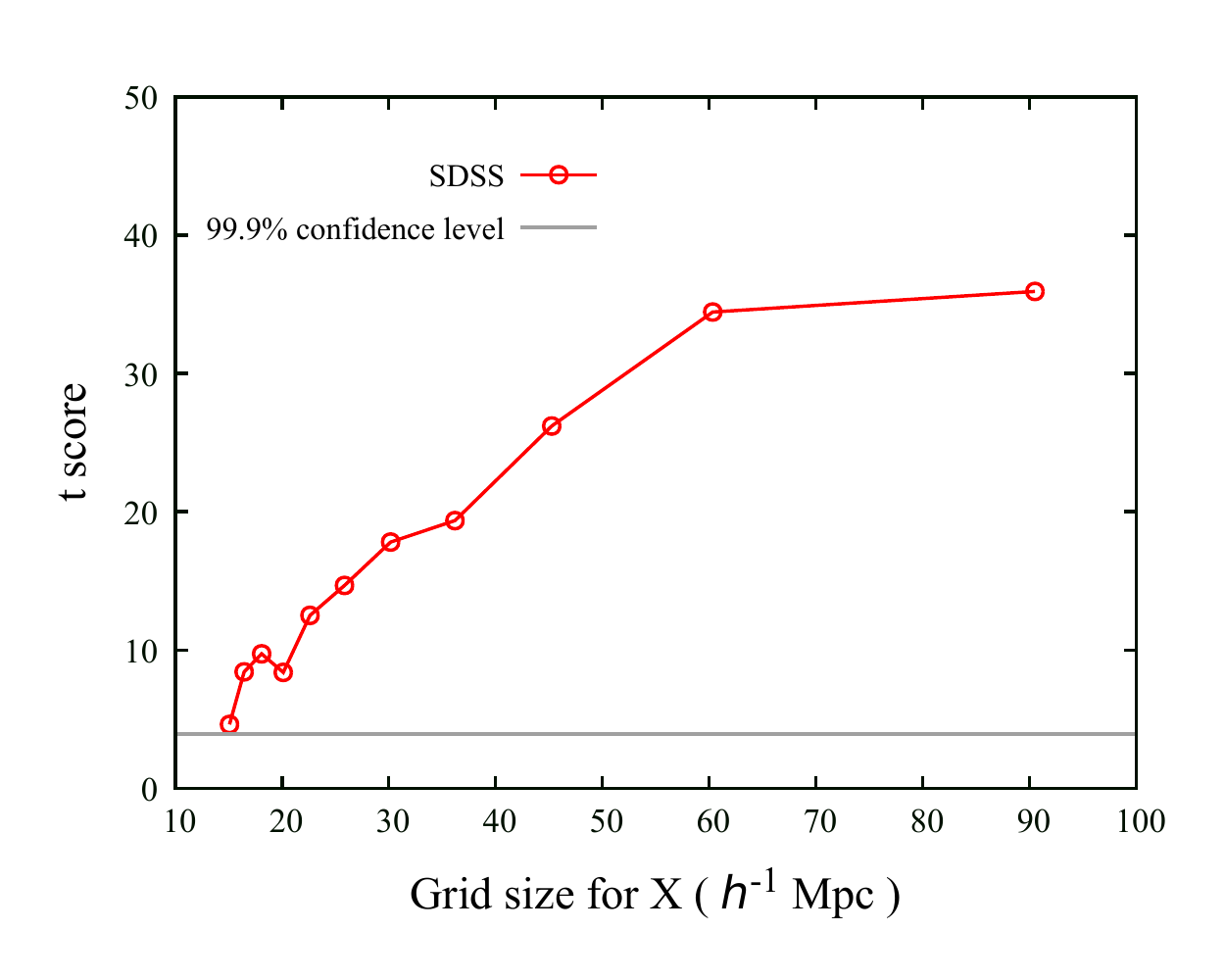}}} \\ 
\resizebox{7.5 cm}{!}{\rotatebox{0}{\includegraphics{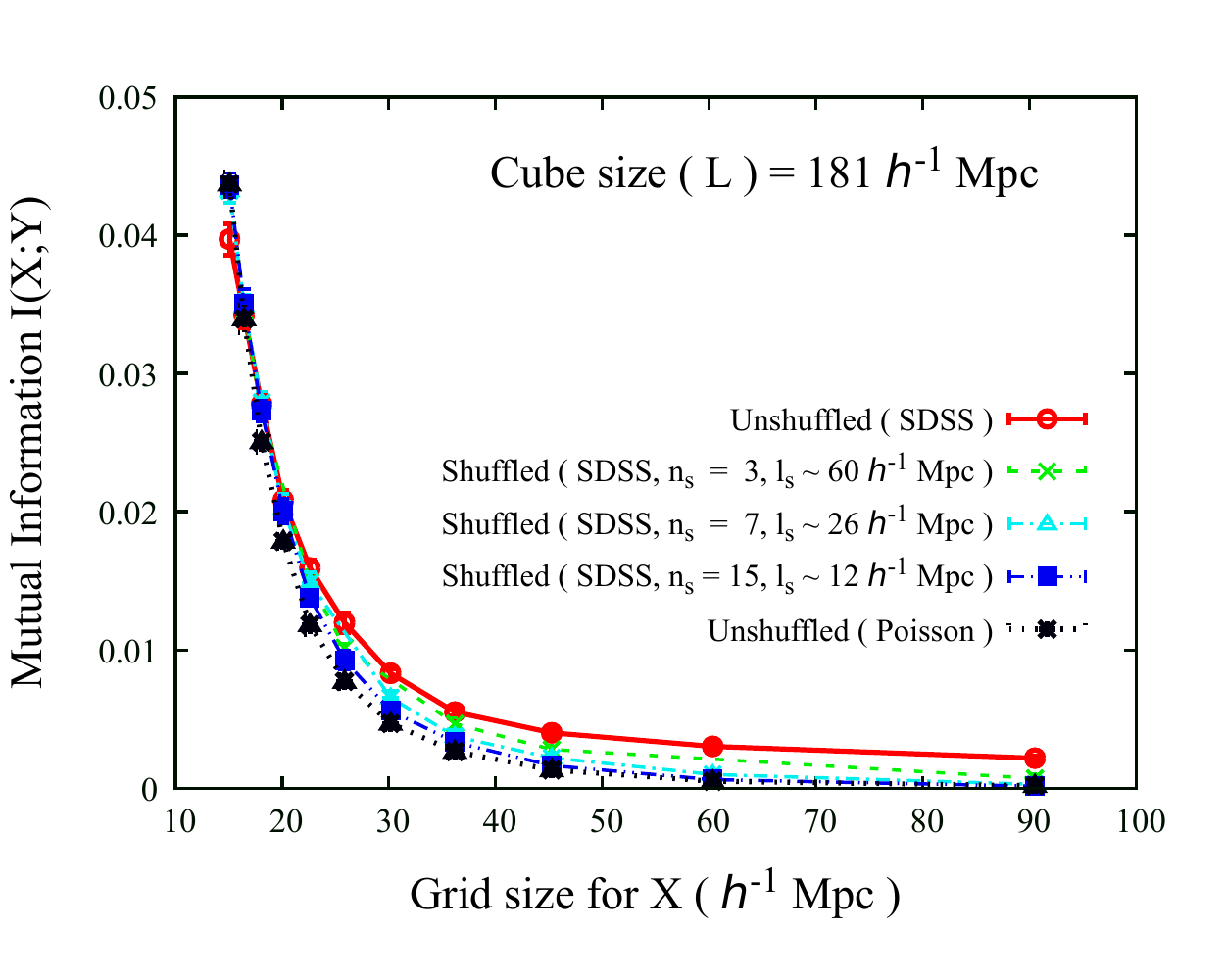}}} \hspace{0.1 cm}
\resizebox{7.5 cm}{!}{\rotatebox{0}{\includegraphics{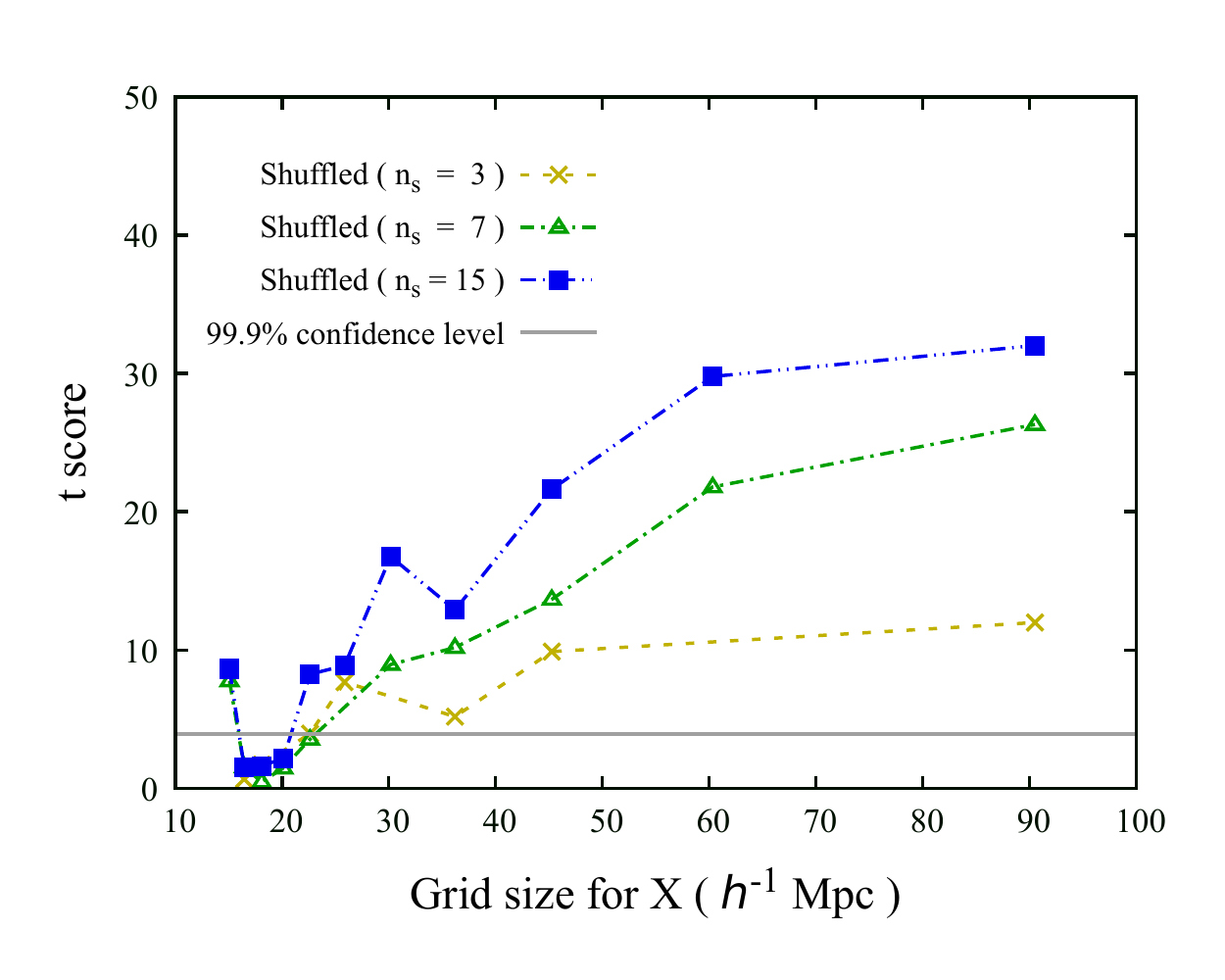}}} \\
\caption{The top left and bottom left panels of this figure show the
  mutual information between colour and environment of spirals as a
  function of length scales. The top left panel shows the effect of
  randomization, and the bottom left panel shows the effect of
  shuffling on the mutual information. The $1-\sigma$ error bars for
  the SDSS data points are obtained by jackknife resampling. It is
  obtained for the randomized and shuffled data using ten different
  realizations in each case. The top right and the bottom right panels
  show the t-score for the differences in the mutual information
  caused by randomization and shuffling, respectively. The threshold
  value of the t-score corresponding to $99.9\%$ confidence level are
  shown together with a horizontal line in these two panels.}
\label{fig:mirb}
\end{figure*}

We measure the mutual information between the colour of spiral
galaxies and their large-scale environment as a function of length
scales. The solid red line in the top left panel of \autoref{fig:mirb}
show the mutual information between the colour and environment of the
spiral galaxies in our volume limited sample. The results show that
the correlation between the colour and the environment of spiral
galaxies decreases with the increasing length scales. We observe a
non-zero mutual information between the two variables throughout the
entire length scales considered here. It is crucial to test the
statistical significance of any such non-zero mutual information. A
statistically significant non-zero mutual information on a given
length scale would suggest that the environment on that length scale
have a role in deciding the colours of spiral galaxies.

We construct ten mock Poisson random distributions within an identical
cubic region, each with the same number of points (10456) as there are
total number of red and blue spirals in the actual SDSS data cube. The
original SDSS data cube contains 8522 blue spirals and 1934 red
spirals. We randomly tag 1934 points as red spirals in each mock
dataset, and the remaining data points are tagged as blue spirals. We
carry out an analysis with these mock datasets in the same way. The
mutual information between the colour and environment of spiral
galaxies in these mock random datasets are shown in the top left panel
of \autoref{fig:mirb} with a dotted black line.  The mutual
information in the actual SDSS dataset is higher than the mock random
dataset at most scales. It may be noted that the mutual information in
the random dataset is higher than in the actual dataset at the
smallest grid size. It indicates the dominance of the Poisson noise on
such length scales. The shot noise decreases with the increasing voxel
sizes, and the differences in the mutual information in the actual and
random data sets are evident at larger length scales.

We further randomize the colour tags of spiral galaxies in the actual
SDSS data without affecting their spatial distributions. The numbers
of red and blue spirals in the randomized datasets are identical to
the original SDSS data. We generate ten such new distributions and
carry out an analysis in the same way. The results are shown in the
top left panel of \autoref{fig:mirb} where we find that mutual
information in the randomized data sets is nearly identical to that
with the mock Poisson random distribution. The randomization of the
colour tags reduces the mutual information at nearly all length
scales. We use a t-test to assess the difference between the mutual
information in the original and the randomized data sets. The
resulting t-score at each length scale is shown in the top right panel
of \autoref{fig:mirb}. The threshold value of the t-score
corresponding to $99.9\%$ confidence level is shown together with a
horizontal line in the same panel. We find that the differences
between the mutual information in the original and randomized data are
statistically significant at $99.9\%$ confidence level at all length
scales probed.

\begin{table*}{}
\caption{This table shows the $t$ score and the associated $p$ value
  at each length scale when we compare the mutual information in the
  actual SDSS data and the SDSS data with randomized classification.}
\label{tab:ttest1}
\center
\begin{tabular}{ccc}
\hline
 Grid size & $t$ score & $p$ value \\
 ($\hmpc$) &  & \\

\hline

$15.08$ & $ 4.619$ & $1.06 \times 10^{-4}$\\
$16.45$ & $ 8.419$ & $5.87 \times 10^{-8}$\\
$18.10$ & $ 9.733$ & $6.76 \times 10^{-9}$\\
$20.11$ & $ 8.385$ & $6.23 \times 10^{-8}$\\
$22.62$ & $12.513$ & $1.28 \times 10^{-10}$\\
$25.86$ & $14.679$ & $9.24 \times 10^{-12}$\\
$30.17$ & $17.814$ & $3.51 \times 10^{-13}$\\
$36.20$ & $19.359$ & $8.44 \times 10^{-14}$\\
$45.25$ & $26.193$ & $4.37 \times 10^{-16}$\\
$60.33$ & $34.434$ & $3.49 \times 10^{-18}$\\
$90.50$ & $35.924$ & $1.64 \times 10^{-18}$\\
\hline
\end{tabular}
\end{table*} 

\begin{table*}{}
\caption{This table shows the $t$ score and the associated $p$ value
  at each length scale when we compare the mutual information between
  actual SDSS data and its shuffled realizations for different
  shuffling lengths. The grid size for each $n_s$ is chosen so that
  the shuffling length is not equal or an integral multiple of the
  grid size.}
\label{tab:ttest2}
\begin{tabular}{cccccccc}
\hline
 Grid size   & \multicolumn{2}{c}{$n_s = 3$}  & \multicolumn{2}{c}{$n_s = 7$}  & \multicolumn{2}{c}{$n_s = 15$}\\
 ( $\hmpc$ ) & $t$ score & $p$ value & $t$ score & $p$ value & $t$ score & $p$ value 	  \\
\hline

$15.08$	&	-		&			-				&	$ 7.777$		&	$1.82 \times 10^{-7}$		&	$ 8.649$		&	$3.96 \times 10^{-8}$\\
$16.45$	& $0.676$	&	$2.53 \times 10^{-1}$	&	$ 1.566$		&	$6.73 \times 10^{-2}$		&	$ 1.541$		&	$7.02 \times 10^{-2}$\\
$18.10$	& $1.722$	&	$5.10 \times 10^{-2}$	&	$ 0.468$		&	$3.22 \times 10^{-1}$		&	$ 1.617$		&	$6.16 \times 10^{-2}$\\
$20.11$	&	-		&			-				&	$ 1.473$		&	$7.88 \times 10^{-2}$		&	$ 2.161$		&	$2.22 \times 10^{-2}$\\
$22.62$	& $3.962$	&	$4.56 \times 10^{-4}$	&	$ 3.558$		&	$1.12 \times 10^{-3}$		&	$ 8.250$		&	$7.88 \times 10^{-8}$\\
$25.86$	& $7.691$	&	$2.13 \times 10^{-7}$	&		-			&			-					&	$ 8.878$		&	$2.70 \times 10^{-8}$\\
$30.17$	&	-		&			-				&	$ 8.954$		&	$2.37 \times 10^{-8}$		&	$16.770$		&	$9.86 \times 10^{-13}$\\
$36.20$	& $5.185$	&	$3.11 \times 10^{-5}$	&	$10.181$		&	$3.38 \times 10^{-9}$		&	$12.897$		&	$7.83 \times 10^{-11}$\\
$45.25$	& $9.884$	&	$5.33 \times 10^{-9}$	&	$13.659$		&	$3.05 \times 10^{-11}$		&	$21.619$		&	$1.25 \times 10^{-14}$\\
$60.33$	&	-		&			-				&	$21.800$		&	$1.08 \times 10^{-14}$		&	$29.782$		&	$4.55 \times 10^{-17}$\\
$90.50$	& $11.997$	&	$2.53 \times 10^{-10}$	&	$26.307$		&	$4.05 \times 10^{-16}$		&	$32.021$		&	$1.26 \times 10^{-17}$\\
\hline
\end{tabular}
\end{table*}

\subsection{Effects of shuffling on the mutual information}

We divide the SDSS data cube into a number of subcubes and shuffle
them around many times following the method described in
\autoref{sec:shuffle}. The mutual information between colour and
environments of the spiral galaxies before and after shuffling the
spatial distributions are shown in the bottom left panel of
\autoref{fig:mirb}. The results for the mock Poisson samples are also
shown together in the same panel. We find that the mutual information
between colour and environment reduces after shuffling at nearly all
length scales. Shuffling the data is expected to destroy any spatial
coherence in the galaxy distribution beyond the size of the
subcubes. Decreasing the size of the subcubes causes a larger
reduction in the mutual information as this destroys more coherent
patterns at smaller length scales. Shuffling the data with subcubes of
size $12 \hmpc$ causes the mutual information in the shuffled data to
nearly coincide with that from the mock Poisson distributions. It
indicates that the mutual information between the colour and the
large-scale environment would be completely erased when the data is
shuffled with subcubes smaller than $12 \hmpc$.  

We test the statistical significance of the differences in the mutual
information in the original SDSS data and its shuffled realizations
using a t-test. The resulting t-scores at different length scales are
shown in the bottom right panel of \autoref{fig:mirb}. The mutual
information is dominated by the Poisson noise at smaller length
scales. The mutual information in both the shuffled SDSS data and the
original SDSS data are quite similar to unshuffled mock Poisson
datasets at these length scales. It is clear that the differences are
statistically significant at $99.9\%$ confidence level throughout the
entire length scale above $20 \hmpc$. Such statistically significant
reduction in the mutual information due to shuffling indicates that
the observed correlations between colour and environment of spirals on
large-scales are physical.

\section{Conclusions}

We study the distributions of different physical properties like
stellar mass, star formation rate, stellar age of the red and blue
spiral galaxies in a volume limited sample from the SDSS. We compare
the distributions of the red and blue spirals using a
Kolmogorov-Smirnov test and find that the differences are
statistically significant at $99.9\%$ confidence level. We also
compare the distributions of the local density of the red and blue
spirals. Our results show that the red spirals inhabit relatively
denser regions compared to the blue spirals and the local environments
of the red and blue spirals differ statistically at $99.9\%$
confidence level. The galaxy properties are known to be strongly
correlated with their environments. So the local environments of the
spirals must have some roles in transforming their colour and deciding
their physical properties. However, it is essential to test if the
differences in the colour and other physical properties of the spirals
are entirely due to the differences in their local environments alone.
We separately compare the physical properties of the red and blue
spirals in the low-density and high-density regions of our sample and
find that the differences are statistically significant at $99.9\%$
confidence level in both types of environments. It implies that the
local density of environment alone is not sufficient to explain the
differences in the physical properties of the red and blue spirals.
These results are in good agreement with a number of earlier works
\citep{masters10, evans}.

Many earlier studies point out the role of higher density environments
in the formation of red spirals. The environmental processes such as
tidal interactions, minor mergers, thermal evaporation, ram pressure
stripping, galaxy harassment and strangulation may initiate quenching
of star formation in spiral galaxies in such environments. Other
mechanism such as halo quenching, morphological quenching and bar
quenching may also have some role in such transformation. These
mechanisms are either driven by their local environment or triggered
by the internal processes within the galaxies. None of these
environmetal processes or quenching mechanisms can explain the origin
of the red spirals alone. Rather, combinations of these mechanisms is
required to produce the present day population of red spirals
\citep{fraser18}. Even then there will be a large uncertainity in the
evolutionary pathways leading to the red spirals.

In the present work, we test if the colour of the spiral galaxies is
affected by their large-scale environment. We measure the mutual
information between the colour and environment of the spiral galaxies
on different length scales. We find a small but non-zero mutual
information between colour and environment throughout the entire
length scale probed. We randomize the colour tags without affecting
the spatial distribution of the spiral galaxies and find that it
decreases the mutual information at each length scale. The differences
between the original and the randomized datasets are statistically
significant at $99.9\%$ confidence level at all length scales. We then
shuffle the spatial distribution of the spiral galaxies retaining
their original colour tags. The shuffling procedure also decreases the
mutual information at each length scale. A larger drop in mutual
information is observed when the data is shuffled at smaller length
scales. We test the statistical significance of the observed
differences between the original and shuffled datasets. The test shows
that the differences are statistically significant at $99.9\%$
confidence level for nearly the entire length scale.

Several observational studies \citep{pandey06, pandey08, scudder12,
  lietzen12, darvish14, filho15, luparello15, pandey17, pandey20a,
  sarkar20, bhattacharjee20} show that the physical association
between galaxy properties and environment extend much beyond the size
of their host halo. In this work, we test if the colour of galaxies at
fixed stellar mass and fixed morphology is sensitive to the
large-scale environment. Our analysis shows that the correlations
between the colour of the spiral galaxies and their large-scale
environments are statistically significant and hence physical. This
implies that the colour of spiral galaxies at fixed stellar mass
exhibit an additional dependence on their large-scale clustering.
Such an additional dependence on the large-scale clustering may arise
due to the assembly bias. The clustering of dark matter halos are
known to depend on their mass \citep{mo96}. But the clustering also
depends on the assembly history of the dark matter halos
\citep{croton07}. The formation time of smaller halos can differ with
their large-scale environment \citep{jung14}. It has been also
suggested that the mass accretion rates of the dark matter halos may
be correlated even at larger distances if they are hosted in the same
large-scale tidal environment \citep{hearin16}. In the halo model, the
galaxy properties are entirely determined by the mass of the host
halo. But the halo occupation distribution is also sensitive to the
large-scale environment of the halo \citep{gao05, gao07}. Zehavi et
al. \citep{zehavi18} analyze the semi-analytic models applied to the
Millennium simulation and find that the central galaxies at lower halo
mass are more likely to be hosted in the early-forming halos whereas
the opposite is true for the satellite galaxies. A recent analysis
\citep{hadzhi21} of the IllustrisTNG simulations \citep{nelson19}
explores the signarures of the assembly bias on the physical
properties of galaxies and their large-scale distributions.

The morphological transformations are very often associated with the
quenching of star formation. However, the existence of the red spirals
suggests that quenching can also occur without any morphological
transformation. There may exist multiple evolutionary pathways from
the blue cloud to the red sequence. We propose one such pathway based
on the results of our analysis. The assembly history of the galaxies
may play an important role besides their local environment. The mass
assembly history of the spiral galaxies may differ significantly
\citep{dokkum13, rodrig16}. Such differences can make some spirals
more susceptible to mild environmental effects than others. A recent
study based on hydrodynamic simulations indicates that the halo
assembly bias may lead to a wide variation in the amount of cold gas
within halos \citep{cui21}. The early-formed halos accumulate large
cold gas fractions which can delay the onset of quenching. On the
other hand, the late-formed halos have a poor cold gas supply. The
galaxies in the late-formed halos thus remain more vulnerable to
quenching. One can explain the origin of the red spirals by
considering a combined role of the assembly bias and the local
environment. The physical properties of the red and blue spirals
differ both at the low-density and high-density regions. This may
partly arise due to the difference in the assembly history of spiral
galaxies across all types of environments. The quenching in spirals in
the high-density regions is primarily driven by different
environmental effects like minor mergers, stripping and
strangulation. In contrast, internal physical processes like mass
quenching, morphological quenching and bar quenching may play a more
important role in quenching the spirals in the low-density
regions. The spiral galaxies in the early formed halos retain large
amount of cold gas that helps them to maintain a blue colour by
delaying the quenching. The spirals in the late-formed halos may be
easily quenched and turn red due to their poor cold gas
supply. However, it is not possible to fully ascertain this from the
current analysis. Further studies are required to understand better
the roles of assembly bias on the formation of the red spirals.

\section*{ACKNOWLEDGEMENT}
We thank an anonymous reviewer for useful comments and suggestions
that helped us to improve the draft. The authors thank the SDSS team
and the Galaxy Zoo team for making the data publicly available. We
also greatly acknowledge the efforts of the citizen scientists in the
Galaxy Zoo and Galaxy Zoo 2 projects who made the detailed visual
morphological classifications of the SDSS galaxies possible.

BP would like to acknowledge financial support from the SERB, DST,
Government of India through the project CRG/2019/001110. BP would also
like to acknowledge IUCAA, Pune, for providing support through
the associateship programme. SS thanks IISER, Tirupati, for providing
support through a postdoctoral fellowship.

Funding for the SDSS and SDSS-II has been provided by the Alfred
P. Sloan Foundation, the Participating Institutions, the National
Science Foundation, the U.S. Department of Energy, the National
Aeronautics and Space Administration, the Japanese Monbukagakusho, the
Max Planck Society, and the Higher Education Funding Council for
England. The SDSS website is http://www.sdss.org/.

The SDSS is managed by the Astrophysical Research Consortium for the
Participating Institutions. The Participating Institutions are the
American Museum of Natural History, Astrophysical Institute Potsdam,
University of Basel, University of Cambridge, Case Western Reserve
University, University of Chicago, Drexel University, Fermilab, the
Institute for Advanced Study, the Japan Participation Group, Johns
Hopkins University, the Joint Institute for Nuclear Astrophysics, the
Kavli Institute for Particle Astrophysics and Cosmology, the Korean
Scientist Group, the Chinese Academy of Sciences (LAMOST), Los Alamos
National Laboratory, the Max-Planck-Institute for Astronomy (MPIA),
the Max-Planck-Institute for Astrophysics (MPA), New Mexico State
University, Ohio State University, University of Pittsburgh,
University of Portsmouth, Princeton University, the United States
Naval Observatory, and the University of Washington.


\begin{thebibliography}{99}
\bibitem{bond96} J.~R. Bond, L. Kofman, \& D. Pogosyan, \nat, \textbf{380}, 603 (1996)
\bibitem{hubble26} E.~P. Hubble, ApJ, \textbf{64}, 321 (1926)
\bibitem{strateva01} I. Strateva, {\v{Z}}. Ivezi{\'c}, G.~R. Knapp,  V.~K. Narayanan, M.~A. Strauss, J.~E. Gunn, R.~H. Lupton, et al., AJ, \textbf{122}, 1861 (2001)
\bibitem{hogg03} D.~W. Hogg, M.~R. Blanton, D.~J. Eisenstein, J.~E. Gunn, D.~J. Schlegel, I. Zehavi,  N.~A. Bahcall, et al., ApJL, \textbf{585}, L5 (2003)
\bibitem{balogh04} M.~L. Balogh, I.~K. Baldry, R. Nichol, C. Miller,  R. Bower, \&  K. Glazebrook, ApJL, \textbf{615}, L101 (2004)
\bibitem{baldry04} I.~K. Baldry, K. Glazebrook, J. Brinkmann,  {\v{Z}}. Ivezi{\'c}, R.~H. Lupton, R.~C. Nichol,\& A.~S. Szalay, ApJ, \textbf{600}, 681 (2004)
\bibitem{kauffmann03a} G. Kauffmann, T.~M. Heckman, S.~D.~M. White,  S. Charlot, C. Tremonti, E.~W. Peng, M. Seibert, et al., MNRAS, \textbf{341}, 54 (2003)
\bibitem{kannappan04}  S.~J. Kannappan, ApJL, \textbf{611}, L89 (2004)
\bibitem{blanton03} M.~R. Blanton, J. Brinkmann, I. Csabai, M. Doi,  D. Eisenstein, M. Fukugita, J.~E. Gunn, et al., AJ, \textbf{125}, 2348 (2003)
\bibitem{schawinski09a} K. Schawinski, C. Lintott, D. Thomas, M. Sarzi, D. Andreescu, S.~P. Bamford, S. Kaviraj, et al., MNRAS, \textbf{396}, 818 (2009)
\bibitem{masters10} K.~L. Masters, M. Mosleh, A.~K. Romer, R.~C. Nichol, S.~P. Bamford, K. Schawinski, C.~J. Lintott, et al., MNRAS, \textbf{405}, 783 (2010)
\bibitem{fraser16} A. Fraser-McKelvie,  M.~J.~I. Brown,  K.~A. Pimbblet, T. Dolley, J.~P. Crossett, N.~J. Bonne,MNRAS, \textbf{462}, L11 (2016)
\bibitem{vanden} S. van den Bergh, 1976, ApJ, \textbf{206}, 883 (1976)
\bibitem{couch98}  W.~J. Couch,  A.~J. Barger, I. Smail, R.~S. Ellis,  R.~M. Sharples, ApJ, \textbf{497}, 188 (1998)  
\bibitem{dressler99} A. Dressler, I. Smail,  B.~M. Poggianti, H. Butcher,  W.~J. Couch, R.~S. Ellis, A. Oemler, ApJS, \textbf{122}, 51 (1999)
\bibitem{poggianti99} B.~M. Poggianti, I. Smail, A. Dressler, W.~J. Couch, A.~J. Barger, H. Butcher, R.~S. Ellis, et al., ApJ, \textbf{518}, 576 (1999)
\bibitem{goto03} T. Goto, C. Yamauchi, Y. Fujita, S. Okamura, M. Sekiguchi, I. Smail, M. Bernardi, et al., MNRAS, \textbf{346}, 601 (2003)
\bibitem{moran06} S.~M. Moran, R.~S. Ellis, T. Treu, S. Salim,  R.~M. Rich, G.~P. Smith, J.-P. Kneib, ApJL, \textbf{641}, L97 (2006)
\bibitem{wolf09} C. Wolf, A. Arag{\'o}n-Salamanca, M. Balogh, M. Barden,  E.~F. Bell, M.~E. Gray, C.~Y. Peng, et al., MNRAS, \textbf{393}, 1302 (2009)
\bibitem{gallazzi09} A. Gallazzi, E.~F. Bell, C. Wolf, M.~E. Gray, C. Papovich, M. Barden, C.~Y. Peng, et al., ApJ, \textbf{690}, 1883 (2009)
\bibitem{bamford09}  S.~P. Bamford,  R.~C. Nichol,  I.~K. Baldry, K. Land,  C.~J. Lintott, K. Schawinski, A. Slosar, et al., MNRAS, \textbf{393}, 1324 (2009)
\bibitem{skibba09} R.~A. Skibba, S.~P. Bamford, R.~C. Nichol, C.~J. Lintott, D. Andreescu, E.~M. Edmondson, P. Murray, et al., MNRAS, \textbf{399}, 966 (2009)
\bibitem{cowie77} L.~L. Cowie, A. Songaila, Nature, \textbf{266}, 501 (1977)
\bibitem{gunn72} J.~E. Gunn, \& J.~R. Gott, ApJ, \textbf{176}, 1 (1972)
\bibitem{moore96} B. Moore, N. Katz, G. Lake, A. Dressler, \& A. Oemler, Nature, \textbf{379}, 613 (1996)
\bibitem{moore98} B. Moore, G. Lake, \& N. Katz, ApJ, \textbf{495}, 139 (1998) 
\bibitem{larson80} R.~B. Larson, B.~M. Tinsley, \& C.~N. Caldwell, ApJ, \textbf{237}, 692 (1980)
\bibitem{balogh00} M.~L. Balogh, J.~F. Navarro, \& S.~L. Morris, ApJ, \textbf{540}, 113 (2000)
\bibitem{kawata08} D. Kawata, \& J.~S. Mulchaey, ApJL, \textbf{672}, L103 (2008)
\bibitem{birnboim03} Y. Birnboim, \& A. Dekel, MNRAS, \textbf{345}, 349 (2003)
\bibitem{dekel06} A. Dekel, \&  Y. Birnboim, MNRAS, \textbf{368}, 2 (2006)
\bibitem{keres05} D. Kere{\v{s}} , N. Katz, D.~H. Weinberg, \& R. Dav{\'e} , MNRAS, \textbf{363}, 2 (2005)
\bibitem{gabor10} J.~M. Gabor, R. Dav{\'e} , K. Finlator, \& B.~D. Oppenheimer, MNRAS, \textbf{407}, 749 (2010)
\bibitem{combes81} F. Combes, \& R.~H. Sanders, A\&A, \textbf{96}, 164 (1981)
\bibitem{martig09} M. Martig, F. Bournaud, R. Teyssier, \& A. Dekel, ApJ, \textbf{707}, 250 (2009)
\bibitem{peng20}  Y.-. jie Peng ., A. Renzini, MNRAS, \textbf{491}, L51 (2020)  
\bibitem{driver07} S.~P. Driver, C.~C. Popescu,  R.~J. Tuffs, J. Liske,  A.~W. Graham, P.~D. Allen, R. de Propris, MNRAS, \textbf{379}, 1022 (2007)
\bibitem{mahajan09} S. Mahajan, S. Raychaudhury, MNRAS, \textbf{400}, 687 (2009)
\bibitem{cortese12} L. Cortese, A\&A, \textbf{543}, A132 (2012)
\bibitem{bundy06} K. Bundy, R.~S. Ellis,  C.~J. Conselice, J.~E. Taylor, M.~C. Cooper, C.~N.~A. Willmer, B.~J. Weiner, et al., ApJ, \textbf{651}, 120 (2006)
\bibitem{cooper06}  M.~C. Cooper, J.~A. Newman, D.~J. Croton, B.~J. Weiner, C.~N.~A Willmer., B.~F. Gerke, D.~S. Madgwick, et al., MNRAS, \textbf{370}, 198 (2006)
\bibitem{cooper09}  M.~C. Cooper, A. Gallazzi, J.~A. Newman, R. Yan, MNRAS, \textbf{402}, 1942 (2009)
\bibitem{tojeiro13} R. Tojeiro, K.~L. Masters, J. Richards, W.~J. Percival, S.~P. Bamford, C. Maraston,  R.~C. Nichol, et al., MNRAS, \textbf{432}, 359 (2013)
\bibitem{mahajan20} S. Mahajan,  K.~K. Gupta, R. Rana, M.~J.~I. Brown, S. Phillipps, J. Bland-Hawthorn, M.~N. Bremer, et al., MNRAS, \textbf{491}, 398
\bibitem{driver11}  S.~P. Driver, D.~T. Hill,  L.~S Kelvin.,  A.~S.~G. Robotham, J. Liske, P. Norberg,  I.~K. Baldry, et al., MNRAS, \textbf{413}, 971 (2011)
\bibitem{hao19} C.-N. Hao, Y. Shi, Y. Chen, X. Xia, Q. Gu, R. Guo, X. Yu, et al., ApJL, \textbf{883}, L36 (2019)
\bibitem{guo20} R. Guo, C.-N. Hao, X. Xia, Y. Shi, Y. Chen, S. Li, Q. Gu, ApJ, \textbf{897}, 162 (2020)
\bibitem{fraser18} A. Fraser-McKelvie, M.~J.~I. Brown, K. Pimbblet, T. Dolley, N.~J. Bonne, MNRAS, \textbf{474}, 1909 (2018)
\bibitem{evans}  F.~A. Evans, L.~C. Parker, I.~D. Roberts, MNRAS, \textbf{476}, 5284 (2018)
\bibitem{hubble36}  E.P. Hubble, The Realm of the Nebulae (Oxford University Press: Oxford), 79 (1936)
\bibitem{dressler80}  A. Dressler, \apj, \textbf{236}, 351 (1980)
\bibitem{postman84} M. Postman, \& M.~J. Geller, ApJ, \textbf{281}, 95 (1984)
\bibitem{lewis02} I. Lewis, M. Balogh, R. De Propris, W. Couch, R. Bower, A. Offer, J. Bland-Hawthorn, et al., MNRAS, \textbf{334}, 673 (2002)
\bibitem{gomez03} P.~L. G{\'o}mez , R.~C. Nichol, C.~J. Miller, M.~L. Balogh, T. Goto, A.~I. Zabludoff, A.~K. Romer, et al., ApJ, \textbf{584}, 210 (2003)
\bibitem{kauffmann04}  G. Kauffmann, S.~D.~M. White, T.~M. Heckman, et al., \mnras, \textbf{353}, 713 (2004)
\bibitem{hahn07}  O. Hahn, C. Porciani, C.~M. Carollo, \& A. Dekel, \mnras, \textbf{375}, 489 (2007)
\bibitem{gao05} L. Gao, V. Springel, S.~D.~M. White, MNRAS, \textbf{363}, L66 (2005)
\bibitem{wechsler06}  R.~H. Wechsler, A.~R. Zentner,  J.~S. Bullock,  A.~V. Kravtsov, B. Allgood, ApJ, \textbf{652}, 71 (2006)
\bibitem{gao07} L. Gao, \& S.~D.~M. White, MNRAS, \textbf{377}, L5 (2007)
\bibitem{croton07} D.~J. Croton, L. Gao, \& S.~D.~M. White, MNRAS, \textbf{374}, 1303 (2007)
\bibitem{musso18} M. Musso, C. Cadiou, C. Pichon, S. Codis, K. Kraljic, \& Y. Dubois, MNRAS, \textbf{476}, 4877 (2018)
\bibitem{vakili19} M. Vakili, \& C. Hahn, ApJ, \textbf{872}, 115 (2019)
\bibitem{dalal08} N. Dalal, M. White,  J.~R. Bond, A. Shirokov, ApJ, \textbf{687}, 12 (2008)
\bibitem{hahn09} O. Hahn, C. Porciani, A. Dekel, C.~M. Carollo, MNRAS, \textbf{398}, 1742 (2009)
\bibitem{zentner14} A.~R.  Zentner, A.~P.  Hearin, F.~C. van den Bosch, MNRAS, \textbf{443}, 3044 (2014)
\bibitem{mao18} Y.-Y.  Mao, A.~R. Zentner, R.~H. Wechsler, MNRAS, \textbf{474}, 5143 (2018)
\bibitem{zehavi11} I. Zehavi, et al., ApJ, \textbf{736}, 59 (2011)
\bibitem{yan13} H. Yan, Z. Fan, S.~D.~M. White, MNRAS, \textbf{430}, 3432 (2013)
\bibitem{paranjape15} A. Paranjape, K. Kova{\v{c}}, W.~G. Hartley, I. Pahwa, MNRAS, \textbf{454}, 3030 (2015)
\bibitem{lin16} Y.-T. Lin , et al., ApJ, \textbf{819}, 119 (2016)
\bibitem{sin17} L.~P.~T. Sin,  S.~J. Lilly, B.~M.~B. Henriques, MNRAS, \textbf{471}, 1192 (2017)
\bibitem{alam19} S. Alam, Y. Zu, J.~A. Peacock, R. Mandelbaum, MNRAS, \textbf{483}, 4501 (2019)
\bibitem{miyatake16} H. Miyatake, S. More, M. Takada, D.~N. Spergel, R. Mandelbaum, E.~S. Rykoff, \& E. Rozo, PhRvL, \textbf{116}, 041301 (2016)
\bibitem{montero17} A.~D. Montero-Dorta, E. P{\'e}rez, F. Prada, S. Rodr{\'\i}guez-Torres , G. Favole, A. Klypin, R. Cid Fernandes, et al., ApJL, \textbf{848}, L2 (2017)
\bibitem{kerscher18} M. Kerscher, A\&A, \textbf{615}, A109 (2018)
\bibitem{pandey06} B. Pandey, \&  S. Bharadwaj, \mnras, \textbf{372}, 827 (2006)
\bibitem{pandey08}  B. Pandey, \& S. Bharadwaj, \mnras, \textbf{387}, 767 (2008)
\bibitem{scudder12}  J.~M. Scudder, S.~L. Ellison, \& Mendel, J.~T., \mnras, \textbf{423}, 2690 (2012)
\bibitem{lietzen12} H. Lietzen, E. Tempel, P. Hein{\"a}m{\"a}ki, P. Nurmi, M. Einasto, E. Saar, A\&A, \textbf{545}, A104 (2012)
\bibitem{darvish14}  B. Ellison, D. Sobral, B. Mobasher, et al., \apj, \textbf{796}, 51 (2014)
\bibitem{filho15} M.~E. Filho, J. S{\'a}nchez Almeida, C.Mu{\~n}oz-Tu{\~n}{\'o}n, et al., \apj, \textbf{802}, 82 (2015)
\bibitem{luparello15} H.~E. Luparello, M. Lares, D. Paz, et al., \mnras, \textbf{448}, 1483 (2015)
\bibitem{pandey17} B. Pandey, \& S. Sarkar , MNRAS, \textbf{467}, L6 (2017)
\bibitem{pandey20a} B. Pandey, \& S. Sarkar, MNRAS, \textbf{498}, 6069 (2020)
\bibitem{sarkar20} S. Sarkar, \& B. Pandey , MNRAS, \textbf{497}, 4077 (2020)
\bibitem{bhattacharjee20} S. Bhattacharjee, B. Pandey, \& S. Sarkar, JCAP, \textbf{2020}, 039 (2020)
\bibitem{lintott08} C. J. Lintott,  et al,  MNRAS, \textbf{389}, 1179 (2008)
\bibitem{willett13}  K.~W. Willett, C.~J. Lintott, S.~P. Bamford,  K.~L. Masters,  B.~D. Simmons,  K.~R.~V. Casteels,  E.~M. Edmondson, et al., MNRAS, \textbf{435}, 2835 (2013)
\bibitem{ahumada20} R. Ahumada, C. Allende Prieto, A. Almeida, F. Anders, S.~F. Anderson, B.~H. Andrews, B. Anguiano, et al., ApJS, \textbf{249}, 3 (2020)  
\bibitem{york00}  D.~G. York, et al., \aj, \textbf{120}, 1579 (2000)
\bibitem{gunn98} J.~E. Gunn, M. Carr, C. Rockosi, M. Sekiguchi, K. Berry, B. Elms,  E. de Haas, et al., AJ, \textbf{116}, 3040 (1998)
\bibitem{gunn06} J.~E. Gunn, W.~A. Siegmund, E.~J. Mannery, R.~E. Owen, C.~L. Hull, R.~F. Leger, L.~N. Carey, et al., AJ, \textbf{131}, 2332 (2006)
\bibitem{strauss02} M.~A. Strauss, D.~H. Weinberg, R.~H. Lupton, V.~K. Narayanan, J. Annis, M. Bernardi, M. Blanton, et al., AJ, \textbf{124}, 1810 (2002)
\bibitem{bruzual93} A.~G. Bruzual, S. Charlot, ApJ, \textbf{405}, 538 (2004)
\bibitem{brinch04} J. Brinchmann, et al.,  MNRAS, \textbf{351}, 1151 (2004)
\bibitem{conroy09} C. Conroy, J.~E. Gunn, M. White, ApJ, 699, 486 (2009)
\bibitem{sarzi06} M.Sarzi, et al., MNRAS, \textbf{366}, 1151 (2006)
\bibitem{Capp04} M. Cappellari, E. Emsellem,  PASP, \textbf{116}, 138 (2004)
\bibitem{maras11} C. Maraston, G. Str{\"o}mb{\"a}ck, MNRAS, \textbf{418}, 2785 (2011)
\bibitem{thomas11} D. Thomas, C. Maraston, J. Johansson, MNRAS, \textbf{412}, 2183 (2011)
\bibitem{lintott11} C. J. Lintott,  et al,  MNRAS, \textbf{410}, 166 (2011)
\bibitem{planck18} Planck Collaboration, N. Aghanim, Y. Akrami, M. Ashdown, J. Aumont, C. Baccigalupi, M. Ballardini, et al., A\&A, \textbf{641}, A6 (2018)  
\bibitem{pandey20b}  B. Pandey,MNRAS, \textbf{499}, L31 (2020)
\bibitem{zadeh} L. A. Zadeh,Fuzzy sets. Information and Control, \textbf{8}, 338 (1965)
\bibitem{casertano85} S. Casertano, P. Hut,  ApJ, \textbf{298}, 80 (1985)    
\bibitem{mo96} H.~J. Mo,  S.~D.~M. White, MNRAS, \textbf{282}, 347 (1996)
\bibitem{jung14} I. Jung, J. Lee,  S.~K. Yi, ApJ, \textbf{794}, 74 (2014)
\bibitem{hearin16} A.~P. Hearin, P.~S. Behroozi, F.~C. van den Bosch, MNRAS, \textbf{461}, 2135 (2016)
\bibitem{zehavi18} I. Zehavi, S. Contreras, N. Padilla,  N.~J. Smith,  C.~M. Baugh, P. Norberg, ApJ, \textbf{853}, 84 (2018)
\bibitem{hadzhi21} B. Hadzhiyska, S. Liu,  R.~S. Somerville, A. Gabrielpillai, S. Bose, D. Eisenstein, L. Hernquist, MNRAS, 508, 698 (2021)
\bibitem{nelson19} D. Nelson, V. Springel, A. Pillepich, V. Rodriguez-Gomez, P. Torrey, S. Genel, M. Vogelsberger, et al., Computational Astrophysics and Cosmology, \textbf{6}, 2 (2019)
\bibitem{dokkum13} P.~G. van Dokkum, J. Leja, E.~J. Nelson, S. Patel, R.~E. Skelton, I. Momcheva, G. Brammer, et al., ApJL, \textbf{771}, L35 (2013)
\bibitem{rodrig16} V. Rodriguez-Gomez, A. Pillepich, L.~V. Sales, S. Genel, M. Vogelsberger, Q. Zhu, S. Wellons, et al., MNRAS, \textbf{458}, 2371 (2016)
\bibitem{cui21} W. Cui, R. Dav{\'e}, J.~A. Peacock, D. Angl{\'e}s-Alc{\'a}zar, X. Yang, Nature Astronmy, \textbf{5}, 1069 (2021)

\end{thebibliography}
\end{document}